\documentclass[aps,prd,twocolumn,preprintnumbers,nofootinbib,amsmath,amssymb]{revtex4-2}
\usepackage{graphicx}
\usepackage{dcolumn}
\usepackage{bm}
\usepackage{amsmath}
\usepackage{braket}
\usepackage[hidelinks]{hyperref}
\usepackage{epstopdf}
\usepackage{placeins}
\usepackage{multirow}
\usepackage{makecell}
\usepackage{cleveref}
\usepackage{xcolor}

\newcommand{\beq}{\begin{eqnarray}}
\newcommand{\eeq}{\end{eqnarray}}
\newcommand{\beqnn}{\begin{eqnarray*}}
\newcommand{\eeqnn}{\end{eqnarray*}}
\newcommand{\Tr}{\ensuremath{\mathrm{Tr}}}
\newcommand{\SU}{\mathrm{SU}}
\newcommand{\cool}{\mathrm{cool}}
\newcommand{\YM}{\mathrm{YM}}
\newcommand{\clov}{\mathrm{clov}}
\newcommand{\W}{\mathrm{W}}

\renewcommand\thesection{\arabic{section}}

\usepackage[font=small, labelfont=bf, textfont={small,it}]{caption}
\def\spose#1{\hbox to 0pt{#1\hss}}
\def\ltapprox{\mathrel{\spose{\lower 3pt\hbox{$\mathchar"218$}}
\raise 2.0pt\hbox{$\mathchar"13C$}}}

\begin{document}

\title{The $\theta$-dependence of the $\mathrm{SU}(N)$ critical temperature at large $N$}

\author{Claudio Bonanno}
\email{claudio.bonanno@csic.es}
\affiliation{Instituto de F\'isica Te\'orica UAM-CSIC, c/ Nicol\'as Cabrera 13-15, Universidad Aut\'onoma de Madrid, Cantoblanco, E-28049 Madrid, Spain}
\author{Massimo D'Elia}
\email{massimo.delia@unipi.it}
\author{Lorenzo Verzichelli}
\email{lorenzo.verzichelli@unito.it}
\affiliation{Dipartimento di Fisica dell'Universit\`a di Pisa \& \\ INFN Sezione di Pisa, Largo Pontecorvo 3, I-56127 Pisa, Italy}

\date{\today}

\begin{abstract}
We investigate, by means of numerical lattice simulations, the $\theta$-dependence of the critical deconfinement temperature of $\mathrm{SU}(N)$ gauge theories at large $N$: $T_c(\theta) = T_c(0)[1-R\theta^2+O(\theta^4)]$, with $R\sim O(1/N^2)$. We follow two different strategies to determine $R$, one based on the calculation of the latent heat of the transition and on the jump of the topological susceptibility at the $\theta=0$ critical point, the other relying on a direct probe of $T_c(\theta)$ by means of imaginary-$\theta$ Monte Carlo simulations. Our results show that $R$ follows the expected large-$N$ scaling.
\end{abstract}

\maketitle

\section{Introduction}\label{sec:intro}

Non-abelian $\SU(N)$ gauge theories exhibit a non-trivial dependence on the parameter $\theta$, coupling the integer-valued topological charge
\beq\label{eq:topcharge_continuum}
Q = \frac{1}{32 \pi^2} \varepsilon_{\mu\nu\rho\sigma} \int d^4 x \, \Tr\left\{G_{\mu\nu}(x)G_{\rho\sigma}(x)\right\} \in \mathbb{Z}
\eeq
to the standard pure-gauge Yang--Mills action $S_\YM = \frac{1}{2g^2} \int d^4 x \, \Tr\left\{ G_{\mu\nu}(x) G_{\mu\nu}(x)\right\}$, whose behavior in the large-$N$ limit is of particular theoretical and phenomenological interest, due to its relation with the solution of the $\mathrm{U}(1)_{\mathrm{A}}$ puzzle via the Witten--Veneziano mechanism~\cite{tHooft:1973alw,tHooft:1976rip,Witten:1978bc,Witten:1979vv,Veneziano:1979ec,Kawarabayashi:1980dp,Witten:1980sp,DiVecchia:1980yfw}. Among the several intriguing implications of $\theta$-dependence, it is particularly interesting to investigate how the phase structure of Yang--Mills theories is modified in the presence of a non-zero $\theta$~\cite{Unsal:2012zj,Poppitz:2012nz,Anber:2013sga,Kitano:2017jng,Aitken:2018mbb,Chen:2020syd}.

At $\theta=0$, it is well known that $\SU(N)$ pure Yang--Mills theories with $N\ge 3$ undergo a genuine first-order phase transition (for $N=2$, instead, it is of second order~\cite{Fingberg:1992ju}) when crossing the critical temperature $T_c$~\cite{Beinlich:1997ia,Campostrini:1998zd,Lucini:2001ej,Lucini:2002ku,Lucini:2003zr,Lucini:2004yh,Lucini:2005vg,Lucini:2012wq,Borsanyi:2022xml,Lucini:2023irm,Cohen:2023hbq}, which is associated to the spontaneous breaking of the $\mathbb{Z}_N$ center symmetry enjoyed by $S_\YM$, with the Polyakov loop as an order parameter. The critical temperature $T_c$ is required to stay finite in the large-$N$ limit, since the Witten--Veneziano solution to the $\mathrm{U}(1)_{\mathrm{A}}$ problem assumes the confining phase to survive up to $N=\infty$. This behavior has been confirmed by lattice results~\cite{Lucini:2002ku,Lucini:2003zr,Lucini:2004yh,Lucini:2005vg,Lucini:2012wq}.

In the presence of a non-vanishing $\theta$, since $Q$ preserves the center symmetry too, one expects this picture to remain true (at least as long as $\theta < \pi$), with the critical temperature changing as a function of $\theta$. On general grounds, $T_c(\theta)$ is an even function of $\theta$, being $Q$ a $\mathrm{CP}$-odd quantity, whose Taylor expansion around $\theta=0$ is usually parameterized, at leading order, as:
\beq\label{eq:Tc_thetadep_def}
T_c(\theta) = T_c[1-R\theta^2+O(\theta^4)],
\eeq
where $T_c=T_c(0)$ is the $\theta=0$ critical temperature, and where the $O(\theta^2)$ coefficient $R$ represents the curvature of the phase diagram in the $T-\theta$ plane. This picture has been well-verified for $N=3$ from lattice simulations~\cite{DElia:2012pvq,DElia:2013uaf,Otake:2022bcq}, and a numerical computation of $R$, based on the imaginary-$\theta$ dependence of the critical temperature, can be found in~\cite{DElia:2012pvq,DElia:2013uaf}. On the other hand, no lattice investigation of the large-$N$ behavior of $R$ can be found in the literature.

In this respect, in Ref.~\cite{DElia:2012pvq} a prediction for the large-$N$ behavior of $R$ is given, based on the known properties of the $\theta$-dependence of the Yang--Mills free-energy density (here $V$ is the four-dimensional volume)
\beq
\begin{aligned}
f(T,\theta) &\equiv - \frac{1}{V}\log \mathcal{Z}(T,\theta) \\ &\equiv -\frac{1}{V}\log \int [dA] e^{-S_{\YM}+i\theta Q},
\end{aligned}
\eeq
in the broken and the symmetric phases. In particular, the argument of~\cite{DElia:2012pvq} assumes the presence of a first
order transition (which is indeed the case for $N \geq 3$) and relies on the fact that, by definition, at the critical point the free energies in the confined ($c$) and deconfined ($d$) phases are equivalent: $f_c(T_c(\theta),\theta) = f_d(T_c(\theta),\theta)$. Introducing the customary parameterization of the Taylor expansion of $f(T,\theta)$ around $\theta=0$ up to $O(\theta)^2$,
\beq
f(T,\theta) - f(0,\theta) =
\chi(T) \frac{\theta^2}{2} + O(\theta^4),
\eeq
with
\beq
\chi(T) &=& \frac{\braket{Q^2}(T)}{V}\Bigg\vert_{\theta=0}
\eeq
the topological susceptibility, maintaining the equivalence condition as $\theta$ changes, $f_c(T_c(\theta),\theta) = f_d(T_c(\theta),\theta)$, implies:
\beq\label{eq:R_conjecture}
R = \frac{\Delta \chi}{2\Delta \epsilon},
\eeq
where $\Delta \chi=\chi_c-\chi_d$ is the jump of the topological susceptibility at the critical point, while $\Delta \epsilon=\epsilon_d-\epsilon_c$ is the latent heat of the transition, with both quantities computed at $\theta = 0$.

The relation in Eq.~\eqref{eq:R_conjecture} implies that $R\sim O(1/N^2)$, as shown by the following argument~\cite{DElia:2012pvq}. In the confined phase the topological susceptibility approaches a finite large-$N$ limit, as required by the Witten--Veneziano argument, and as it has been extensively checked by lattice simulations~\cite{DelDebbio:2002xa, DElia:2003zne,DelDebbio:2004ns, DelDebbio:2006yuf, Giusti:2007tu, Vicari:2008jw, Panagopoulos:2011rb, Ce:2015qha, Ce:2016awn, Bonati:2015sqt, Bonati:2016tvi, Bonanno:2020hht, Athenodorou:2021qvs}, while higher-order terms in the $\theta$-expansion are expected~\cite{Witten:1998uka}, and have been verified from the lattice~\cite{Bonati:2015sqt,Bonati:2016tvi,Bonanno:2018xtd,Bonanno:2020hht}, to be suppressed as powers of $1/N^2$, or faster. In the deconfined phase, instead, lattice simulations have shown the free energy to approach, as expected, the functional behavior predicted by the Dilute Instanton Gas Approximation (DIGA)~\cite{RevModPhys.53.43,Boccaletti:2020mxu}:
\beq
f(T,\theta) - f(0,\theta) &=& \chi(T) [1-\cos\theta],
\eeq
where $\chi(T) \sim T^{-N}$ is exponentially suppressed in $N$~\cite{Alles:1996nm, Alles:1997qe,Lucini:2004yh,DelDebbio:2004vxo,Bonati:2013tt,Borsanyi:2015cka,Frison:2016vuc,Borsanyi:2016ksw,Kitano:2021jho,Borsanyi:2021gqg}. Putting these facts together, it is thus clear that $\Delta \chi \sim O(N^0)$, and since the latent heat is proportional to the 
number of degrees of freedom, $\Delta \epsilon \sim O(N^2)$, we thus expect, from Eq.~\eqref{eq:R_conjecture}, $R=\Delta\chi/(2 \Delta \epsilon)\sim O(1/N^2)$. However, although this relation between $R$ and the features of the deconfinement transition has been well verified from lattice simulations for $N=3$~\cite{Borsanyi:2022fub}, no direct probe of this prediction at large $N$, nor of the large-$N$ scaling of $R$, can be found in the literature.

Given the lack of lattice studies of the $\theta$-dependence of $T_c$ apart from the $N=3$ case, the goal of this work is to go beyond the state of the art discussed so far, and pursue a direct lattice investigation of the curvature $R$ of the critical temperature at large $N$, with the aim of checking the prediction of~\cite{DElia:2012pvq} in Eq.~\eqref{eq:R_conjecture} in the large-$N$ limit. To do so, we will perform simulations for $N=4$ and $N=6$ and for several imaginary values of $\theta$, and we will determine $R$ from two different and complementary strategies. The first one relies on a $\theta=0$ calculation of $\Delta \chi$ and $\Delta \epsilon$ to determine $R$ via Eq.~\eqref{eq:R_conjecture}. Our second strategy, instead, relies on determining the critical temperature as a function of the imaginary $\theta$, so that $R$ can be directly obtained from its definition in Eq.~\eqref{eq:Tc_thetadep_def}. Finally, we will discuss the large-$N$ behavior of $R$ combining our new results with the pre-existing ones for $N=3$.

This manuscript is organized as follows. In Sec.~\ref{sec:setup} we will present our numerical setup. In Sec.~\ref{sec:res} we will present our numerical results for $R$ for $\SU(4)$ and $\SU(6)$, and we will discuss the large-$N$ limit of the curvature combining our determinations with the $N=3$ one. Finally, in Sec.~\ref{sec:conclu} we will draw our conclusions.

\section{Numerical setup}\label{sec:setup}

In this section we will describe our lattice setup, namely, the adopted discretization of the action, the adopted algorithms used for Monte Carlo (MC) simulations, and the strategy pursued to identify the critical point and calculate the critical temperature and other relevant observables.

\subsection{Lattice action}\label{sec:lat_action}

We discretize the $\SU(N)$ pure-gauge action on a $N_s^3 \times N_t$ lattice with periodic boundary conditions in all directions, where the temperature is given by $T=(a N_t)^{-1}$, with $a$ the lattice spacing. Concerning the $\theta=0$ theory, we adopted the standard Wilson lattice action:
\beq\label{eq:wilson_action}
S_\W[U] = \beta \sum_{x, \mu < \nu} \left( 1 - \frac{1}{N} \Re \, \Tr\left[\Pi_{\mu\nu}(x)\right] \right),
\eeq
where $\beta=2N/g^2$ is the bare coupling and $\Pi_{\mu\nu}(x)=U_\mu(x)U_\nu(x+\hat{\mu})U_\mu^\dagger(x+\hat{\nu})U_\nu^\dagger(x)$ is the product of the $\SU(N)$ gauge link matrices along the plaquette in the $(\mu,\nu)$ plane based on the site $x$ of the lattice.

Concerning the theory at non-zero $\theta$, as it is well known, doing direct MC simulations at real values of $\theta$ is impossible due to the sign problem introduced by the topological term, which is purely imaginary:
\beq\label{eq:theta_dep_action}
S_{\YM} \rightarrow S(\theta) = S_{\YM}  -i\theta Q.
\eeq

A common strategy to deal with this issue, which has been widely employed in lattice simulations targeting the study of $\theta$-dependence~\cite{Bhanot:1984rx,Azcoiti:2002vk,Alles:2007br,Imachi:2006qq,Aoki:2008gv,Panagopoulos:2011rb,
Alles:2014tta,DElia:2012pvq,DElia:2013uaf,DElia:2012ifm,Bonati:2015sqt, Bonati:2016tvi,
Bonati:2018rfg, Bonati:2019kmf,Bonanno:2018xtd,Berni:2019bch,Bonanno:2020hht}, is to perform imaginary-$\theta$ simulations. Assuming analyticity around $\theta=0$, it is possible to perform an analytic continuation towards imaginary values of $\theta$ with the change of variables $\theta_I = i \theta$, so that the action in~\eqref{eq:theta_dep_action} becomes purely real and the sign problem due to the $\theta$-term is avoided.

Concerning the topological term, we define it in terms of the so-called clover discretization, which is the simplest lattice discretization of the topological charge $Q$, defined in Eq.~\eqref{eq:topcharge_continuum}, with a definite parity:
\beq\label{eq:clov_charge}
\begin{aligned}
Q_{\clov} &= \sum_{x} q_\clov(x) \\ &= \frac{1}{2^9 \pi^2} \sum_{\mu\nu\rho\sigma = \pm1}^{\pm4} \varepsilon_{\mu\nu\rho\sigma} \Tr\left[ \Pi_{\mu\nu}(x) \Pi_{\rho\sigma}(x) \right],
\end{aligned}
\eeq
where the Levi--Civita tensor with a negative index is defined as $\varepsilon_{(-\mu)\nu\rho\sigma} = -\varepsilon_{\mu\nu\rho\sigma}$. In the end, our total lattice action looks like:
\beq\label{eq:lat_action_theta_I}
S_L(\theta_I) = S_\W - \theta_L Q_\clov.
\eeq

The discretization of the charge in Eq.~\eqref{eq:clov_charge} is not integer valued, and in particular its value is related to the physical integer one, on average, by a multiplicative renormalization,
\beq\label{eq:renorm_Qclov}
Q_\clov = Z(\beta) Q,
\eeq
which stems from UV fluctuations at the scale of the lattice spacing~\cite{Campostrini:1988cy,Vicari:2008jw}. This, in particular, leads to a multiplicative renormalization of the lattice parameter $\theta_L$ as follows:
\beq
\theta_I = Z(\beta) \theta_L,
\eeq
where $\theta_I$ is the physical imaginary-$\theta$.

Concerning the topological susceptibility, instead, a naive definition based on $Q_\clov$ would suffer also for an additional additive renormalization due to short-distance contact terms~\cite{DiVecchia:1981aev,DElia:2003zne}:
\beq
\frac{\braket{Q_\clov^2}}{V} = Z(\beta)^2 \frac{\braket{Q^2}}{V} + M(\beta).
\eeq
Such additive renormalization would eventually become dominant in the continuum limit compared to the physical signal.

A customary choice to get rid of renormalization effects is to compute the topological charge on smoothened configurations. Smoothing is a procedure to dampen UV fluctuations at the scale of the lattice cut-off $\sim 1/a$ while leaving the relevant IR topological fluctuations intact. Many different smoothing algorithms have been proposed, such as cooling~\cite{Berg:1981nw,Iwasaki:1983bv,Itoh:1984pr,Teper:1985rb,Ilgenfritz:1985dz,Campostrini:1989dh,Alles:2000sc}, gradient flow~\cite{Luscher:2009eq, Luscher:2010iy} or stout smearing~\cite{APE:1987ehd, Morningstar:2003gk}, all giving agreeing results when properly matched to each other~\cite{Alles:2000sc, Bonati:2014tqa, Alexandrou:2015yba}.

In this work we adopted cooling for its simplicity and numerical cheapness. A single step of cooling consists of a lattice sweep where each link is aligned to its local staple (computed always at $\theta=0$), in order to locally minimize the Wilson action. After cooling, $Z\simeq 1$, and the distribution of $Q_\clov^{(\cool)}$ shows sharp peaks very close to integer values. Thus, we assign to each configuration an integer topological charge via the following procedure~\cite{DelDebbio:2002xa}:
\beq\label{eq:rounded_topcharge}
Q = \mathrm{round}\{\alpha \, Q_\clov^{(\cool)}\},
\eeq
where ``$\mathrm{round}\{x\}$'' denotes rounding $x$ to the closest integer, and $\alpha$ is defined as the minimum of
\beq
\braket{(\alpha Q_\clov^{(\cool)} - \mathrm{round}\{\alpha \, Q_\clov^{(\cool)}\})^2}.
\eeq
The typical values we found for $\alpha$ were of the order of $\sim 1.05$--$1.1$.

Since, after smoothing, one also has $M\simeq 0$, the quantity $\chi = \braket{Q^2}/V$ is a proper lattice definition of the topological susceptibility, with a proper scaling towards the continuum limit. Moreover, smoothing allows us to perform a non-perturbative computation of the constant $Z$ needed to renormalize the lattice parameter $\theta_L$:
\beq\label{eq:charge_renorm}
Z = \frac{\braket{Q Q_{\clov}}}{\braket{Q^2}},
\eeq
cf.~Eq.~\eqref{eq:renorm_Qclov}. In all explored cases, we found that the obtained results for $Q$, for $\chi$ and $Z$ were stable after $\sim 10$ cooling steps, thus we always computed the lattice rounded charge after 30 cooling steps.

\subsection{Monte Carlo algorithms}\label{sec:algos}

In most cases, to generate gauge configurations we adopted the standard local over-relaxation (OR)~\cite{Creutz:1987xi} and heat-bath (HB)~\cite{Creutz:1980zw,Kennedy:1985nu} algorithms. More precisely, our single MC updating step is given by 4 lattice sweeps of OR, followed by a lattice sweep of HB, where both local updates have been implemented \emph{\`a la} Cabibbo--Marinari~\cite{Cabibbo:1982zn}, i.e., updating all the $N(N-1)/2$ diagonal $\SU(2)$ subgroups of $\SU(N)$.

Local algorithms for $\SU(N)$ gauge theories are well known to suffer, when approaching the continuum limit, from a severe increase of the autocorrelation time of topological quantities, usually known as ``topological critical slowing down''. Such problem further worsens as $N$ is increased, eventually leading to a complete freezing of the MC evolution of the topological charge density even for moderate coarse lattice spacings when $N$ is sufficiently large~\cite{Alles:1996vn, 
deForcrand:1997yw,Lucini:2001ej,DelDebbio:2002xa,Leinweber:2003sj,
DelDebbio:2004xh,
Luscher:2011kk,
Laio:2015era, Flynn:2015uma,
Bonati:2015vqz,
Bonati:2017woi,Athenodorou:2020ani,Athenodorou:2021qvs}. Since we are interested in measuring the $\theta$-dependence of the critical temperature, topological freezing can potentially introduce sizeable systematic effects in $R$. Moreover, topological freezing makes it very difficult to obtain a reliable estimation of the topological susceptibility at large $N$, which is necessary to accurately check Eq.~\eqref{eq:R_conjecture}. Since, as we will show in the following, for the finest lattice spacings explored at our largest value of $N$, the MC histories of the topological charge are almost frozen, it is clear that in those cases a strategy to mitigate topological freezing is needed to keep possible systematics affecting $R$ under control.

To this end, for those simulations points, we adopted the Parallel Tempering on Boundary Conditions (PTBC) algorithm. This algorithm was first proposed~\cite{Hasenbusch:2017unr} and employed~\cite{Berni:2019bch,Bonanno:2022hmz} in $2d$ $\mathrm{CP}^{N-1}$ models, and later implemented for $4d$ $\SU(N)$ gauge theories too~\cite{Bonanno:2020hht,Bonanno:2022yjr,DasilvaGolan:2023cjw}.

The PTBC algorithm involves the simultaneous simulation of $N_{r}$ copies of the system differing only for the boundary conditions imposed on a small 3-dimensional cubic $L_d^3$ region, orthogonal to a spatial direction of the lattice, referred to as the ``defect''. One of the replicas, say the one labelled by $r = 0$, has standard periodic boundary conditions, while, for the other replicas ($r > 0$), each link crossing the defect gets multiplied in the Wilson action by a factor $c(r)$, interpolating between $c(0) = 1$ and $c(N_r- 1) = 0$. The topological term is instead not altered in any replica.

The last replica ($r = N_r - 1$) has open boundary conditions along the defect, thus suffering for much smaller autocorrelation times of the topological charge~\cite{Luscher:2011kk,Luscher:2012av}. The fast MC evolution of the topological charge achieved in the open replica is then transferred towards the periodic one, since, after each replica has been independently updated via the standard MC step earlier introduced, the swap of each couple of adjacent replicas ($r$, $r+1$) is proposed. Swaps are proposed via a Metropolis test whose acceptance probability is:
\beqnn
\label{eq:PTBC_swap_rate}
p(r, r+1) &=& \min\left\{ 1, \exp\left( -\Delta S \right) \right\},
\eeqnn
\beqnn
\Delta S = && \left[ S_W^{(r)}[U^{(r+1)}] + S_W^{(r+1)}[U^{(r)}] \right] +
\\&&- \left[ S_W^{(r)}[U^{(r)}] + S_W^{(r+1)} [U^{(r+1)}]\right],
\eeqnn
where $S_\W^{(r)}$ denotes the Wilson action in the presence of the boundary conditions of the replica $r$, and $U^{(r)}$ denotes the gauge fields of the replica $r$.

To ensure the efficiency of the swapping procedure, the $c(r)$ coefficients are tuned so that the probability of accepting the swaps during the Metropolis test is approximately constant for each pair of exchanged replicas ($r, r+1$), i.e., $\braket{p(0, 1)} \approx \braket{p(1, 2)} \approx \dots \approx \braket{p(N_r - 2, N_r - 1)}$, which allows a given configuration to perform a random walk among the different replicas with different boundary conditions.

The advantage of using the PTBC algorithm is that it allows to exploit the much smaller autocorrelation times of the topological charge achieved with open boundary conditions, while at the same time bypassing several complications introduced by them. Indeed, this algorithm allows to perform measures directly in the periodic replica, where no issue related to the unphysical contributions of the boundary is present (which would require larger lattices to be kept under control), and where translation invariance is unbroken, so that a notion of global topological charge is preserved. Thus, all observables will always be computed on the periodic lattice.

\subsection{Identification of the critical point}\label{sec:crit_point}

The order parameter of the phase transition is given by the expectation value of the Polyakov loop, which we discretize as follows:
\beq
L = \sum_{\vec{x}} L(\vec{x}) \equiv \frac{1}{N} \frac{1}{N_s^3} \sum_{\vec{x}} \Tr \left\{ \prod_{t = 1}^{N_t} U_0(\vec{x}, t)\right\},
\eeq
where $\vec{x}$ is the space position in each 3-dimensional slice of the lattice at fixed Euclidean time $t$. As usual, we will study the expectation value of its modulus, $\braket{ \vert L \vert}$.

To look for the critical point, we will follow the so-called ``fixed-$N_t$''approach. Namely, for a given lattice, we change the temperature $T$ by changing the bare coupling $\beta$ according to $1/T=a(\beta) N_t$, and we identify he critical point by looking at the behavior of the susceptibility of the Polyakov loop
\beq
\chi_L \equiv N_s^3 \left( \braket{|L|^2} - \braket{|L|}^2 \right),
\eeq
as a function of $\beta$. Indeed, on a finite volume, $\chi_L$ presents a peak of finite width in correspondence of the critical temperature, associated to the critical coupling $\beta_c$.

To determine more precisely such value of $\beta$, we interpolated the values of $\chi_L(\beta)$ computed with direct simulations using the standard ``multihistogram method''~\cite{multi_histogram}. After the multihistogram analysis, we obtain a smooth curve for $\chi_L(\beta)$. We then fitted the peak of the interpolated curve with a Lorentzian function centered around $\beta_c$. The uncertainties on the fit parameters were estimated from $O(1000)$ binned bootstrap resamplings of the original data. A proper systematic uncertainty has been also associated to the variation observed upon varying the fit range around the peak.

To pass from the critical $\beta$ to the critical temperature, we need to set the scale. To this end we used the determinations of the string tension in lattice units $a(\beta)\sqrt{\sigma}$ reported in~\cite{Lucini:2005vg}. In particular, we determined the critical temperature in units of the string tension as follows:
\beq\label{eq:crit_temp}
\frac{T_c}{\sqrt{\sigma}} = \frac{1}{N_t \, a(\beta_c) \, \sqrt{\sigma}},
\eeq

Once the critical point is identified, we will be interested in two observables computed at $\beta_c$ (which will be computed only for $\theta = 0$): the latent heat and the discontinuity of the topological susceptibility. In both cases, we will need to compute differences between expectation values in the confined and deconfined phases. To this end, one needs a clear separation between the two phases, i.e., two clearly distinct peaks in the histogram of $\vert L \vert$, in order to assign a configuration to either a phase or the other. In practice, thus, this requires to work on large enough volumes, and to compute the quantities of interest only considering the gauge configurations with $\vert L \vert$ lying in the range of values that are typical of each phase. More details about this procedure will be given in Sec.~\ref{sec:res}.

Concerning the jump of the topological susceptibility, we computed the discontinuity as follows:
\beq\label{eq:jump_chi}
\frac{\Delta \chi}{T_c^4} = \left(\frac{N_t}{N_s}\right)^3 \left( \braket{Q^2}_c - \braket{Q^2}_d \right)
\eeq
where $Q$ is the rounded charge defined in Eq.~\eqref{eq:rounded_topcharge}, and where $\braket{\mathcal{O}}_{c/d}$ denote the expectation values in the confined ($c$) and deconfined ($d$) phases.

Regarding the latent heat, instead, we computed the jump at the critical temperature of the trace anomaly $(\epsilon - 3p) / T_c^4$. Indeed, this coincides with the discontinuity of the internal energy $\epsilon$, since the pressure $p$ is continuous. The trace anomaly can be computed on the lattice using the thermodynamical identity:
\beq
\frac{\epsilon - 3p}{T^4} =
T \frac{\partial}{\partial T} \left( \frac{p}{T^4}  \right) =
T \frac{\partial}{\partial T} \left( \frac{1}{V T^4} \log(\mathcal{Z}) \right),
\label{trace_anomaly}
\eeq
where $\mathcal{Z} = \int d[U] \exp(-S_\W[U])$ is the partition function. Rewriting Eq.~\eqref{trace_anomaly} in terms of lattice quantities yields:
\beq
\frac{\varepsilon - 3p}{T^4} =
N_t^4 \frac{\partial \beta}{\partial \log[a(\beta)\sqrt{\sigma}]} \frac{\braket{S_\W}}{V\beta}.
\eeq
The latent heat $\Delta \varepsilon$ can, thus, be related to the difference in the expectation values of the mean plaquette between the two phases:
\beq\label{eq:latent_heat}
\frac{\Delta\varepsilon}{T_c^4} = 6 N_t^4\frac{\partial \beta}{\partial \log [a(\beta)\sqrt{\sigma}]}\Bigg\vert_{\beta_c}\left(\braket{U}_c - \braket{U}_d \right),
\eeq
where
\beq
U \equiv \frac{1}{6 N N_s^3 N_t}\sum_{x, \mu>\nu} \Re \Tr \left\{\Pi_{\mu\nu}(x)\right\}.
\eeq

\section{Results}\label{sec:res}

This section is devoted to the discussion of our results for $R$ obtained for $N=4$ and $6$ both from the imaginary-$\theta$ dependence of the critical temperature and from the determination of the latent heat of the transition and of the gap in the topological susceptibility measured at $\theta=0$. In both cases we will discuss both the extrapolation towards the continuum limit, and the extrapolation towards the large-$N$ limit.

\subsection{Results for $\SU(4)$}\label{sec:res_N4}

\begin{figure}[!t]
\centering
\includegraphics[width=0.475\textwidth]{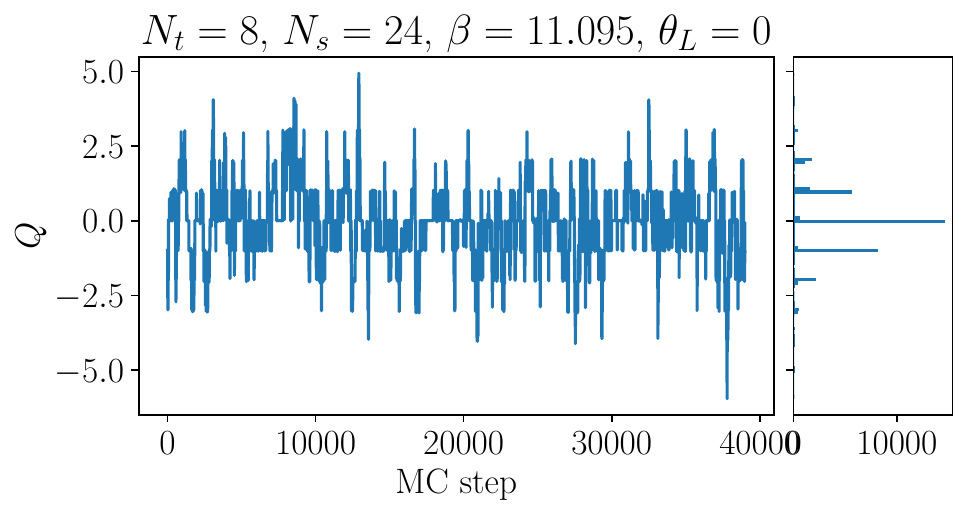}
\includegraphics[width=0.475\textwidth]{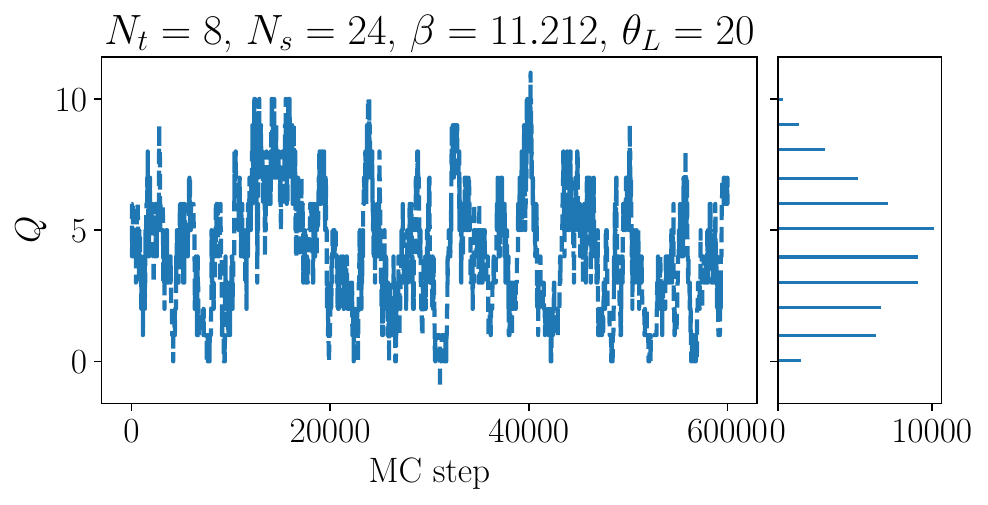}
\caption{Evolution of the topological charge $Q$ defined in Eq.~\eqref{eq:rounded_topcharge} for the largest $\beta$ explored at $\theta_L=0$ and $\theta_L=20$ on a $24^3 \times 8 $ lattice for $N=4$. The horizontal MC time is expressed in terms of the number of single MC steps defined at the beginning of Sec.~\ref{sec:algos}.}
\label{fig:topcharge_evo_N4}
\end{figure}

We start our discussion from the determination of $R$ in the case $N=4$. We considered 3 values of the temporal extent, namely, $N_t=5,6,8$. For each lattice, we performed simulations for a few values of $\theta_L$, and for each value of the imaginary-$\theta$ we performed simulations for several values of $\beta$, in order to identify the critical point. For this value of $N$, we observed a reasonable number of topological fluctuations in all explored points, thus, all simulations were performed with the standard local algorithm, without relying on the parallel tempering. Examples of MC histories of the topological charge are shown in Fig.~\ref{fig:topcharge_evo_N4}

\begin{figure}[t!]
\centering
\includegraphics[width=0.475\textwidth]{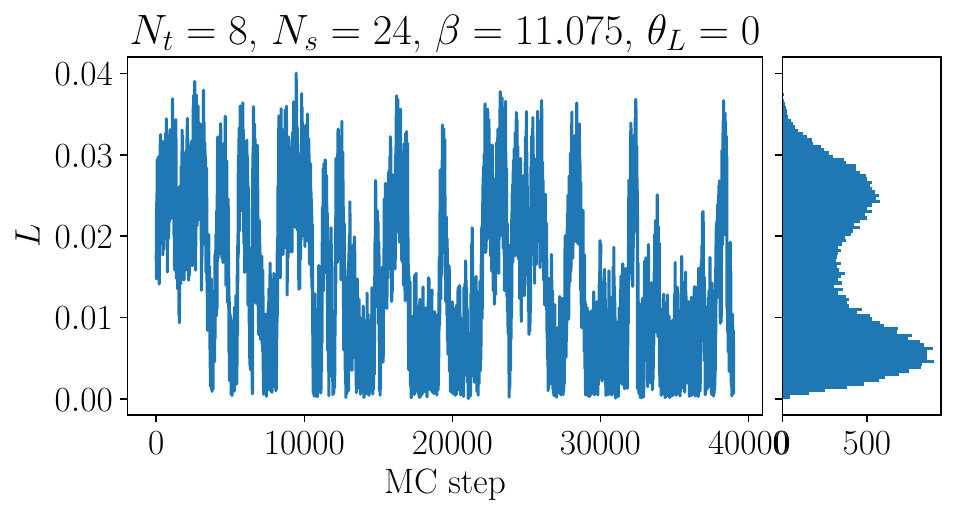}
\includegraphics[width=0.475\textwidth]{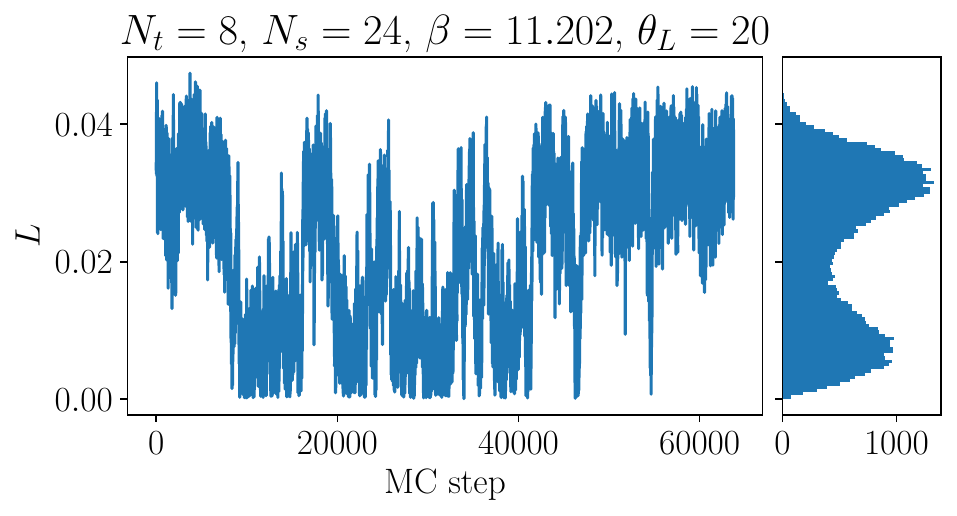}
\caption{Evolution of the Polyakov loop $L$ for $\beta \simeq \beta_c$ and $\theta_L=0, 20$ on a $24^3 \times 8$ lattice for $N=4$. The horizontal MC time is expressed in terms of the number of single MC steps defined at the beginning of Sec.~\ref{sec:algos}.}
\label{fig:poly_evo_N4}
\end{figure}

\begin{figure*}[!htb]
\centering
\includegraphics[width=0.96\textwidth]{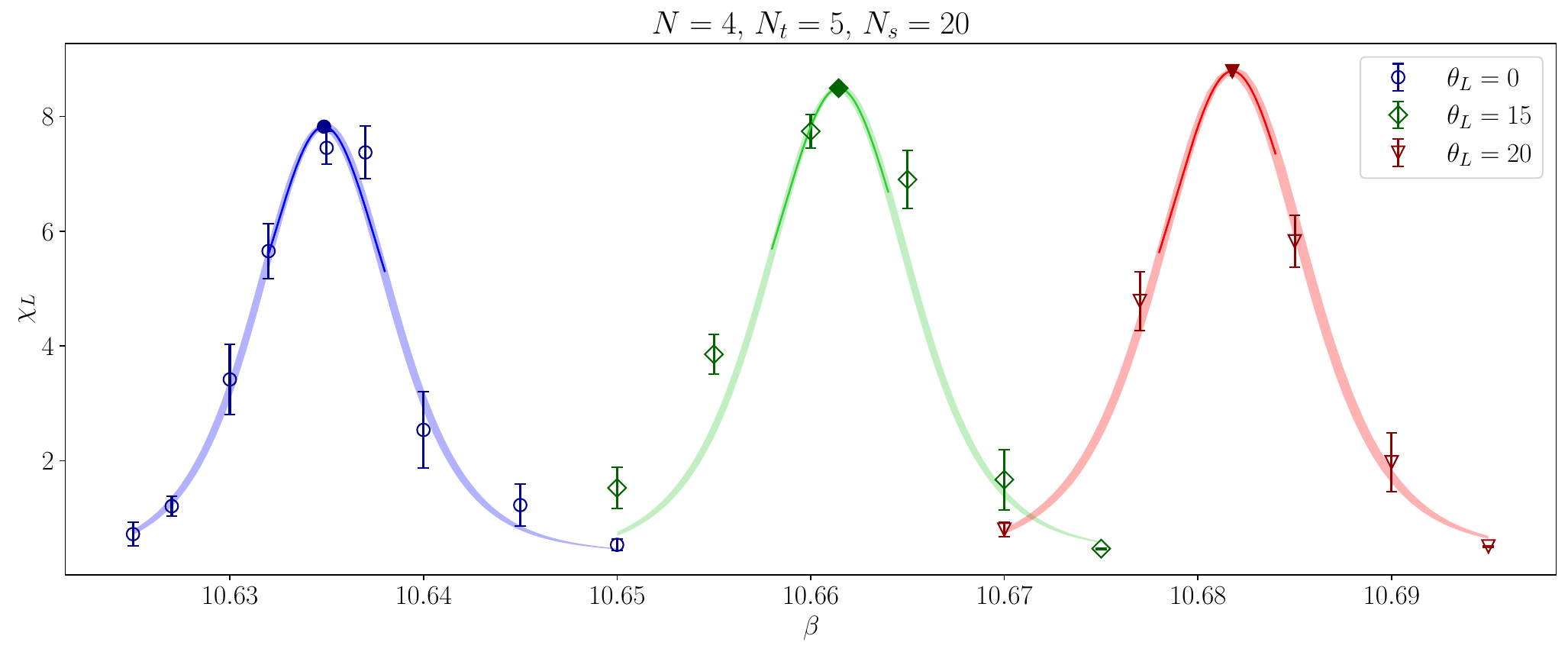}
\caption{Peaks of the Polyakov loop susceptibility $\chi_L(\beta)$ as a function of the bare coupling $\beta$ observed on a $20^3\times5$ lattice for 3 different values of $\theta_L=0,15,20$ and $N=4$. The lighter shaded areas represent the multihstogram interpolations of the MC data with its errors, while the darker solid lines are the results of the best fits of the interpolated data according to a Lorentzian ansatz $f(\beta)=A_1/[1+(\beta-\beta_c)^2/A_2]$. Filled points represent the positions of the critical points.}
\label{fig:chi_L_peaks}
\end{figure*}

For each value of $\theta_L$, we observe that the Monte Carlo histories of the Polyakov loop, when the critical point is approached, exhibit the typical bistability which characterizes the coexistence of the broken and the symmetric phases, cf.~Fig.~\ref{fig:poly_evo_N4}. This, in turn, results in a peak of the Polyakov susceptibility $\chi_L$ as a function of $\beta$ in correspondence of the critical coupling $\beta_c$, which is computed by means of the multihistogram analysis explained in Sec.~\ref{sec:crit_point}. As expected, the peak of $\chi_L(\beta)$, corresponding to the critical $\beta_c$, moves towards higher couplings as $\theta_L$ is increased, see Fig.~\ref{fig:chi_L_peaks}.

All determinations of $\beta_c(\theta_L)$ were done on lattices with aspect ratios $N_s/N_t=4$, except for the simulations performed for $N_t=8$, where we used $N_s/N_t=3$. Since we checked that, for our lattice with $N_t=6$, and for the smallest and largest values of $\theta_L$ explored, the results obtained for $\beta_c(\theta_L)$ with aspect ratios $N_s/N_t=3$ and $N_s/N_t=4$ were perfectly compatible, we conclude that finite size effects are well under control. All determinations of the critical values of $\beta$ found for $N=4$ are reported in Tab.~\ref{tab:SU4_results}.

After determining the critical coupling $\beta_c$ as a function of $\theta_L$, we determined $T_c/\sqrt{\sigma}$ according to Eq.~\eqref{eq:crit_temp}, and the physical imaginary-$\theta$ parameter $\theta_I = Z(\beta_c) \theta_L$ according to the determinations of $Z(\beta_c)$ obtained from Eq.~\eqref{eq:charge_renorm}. We remark that the renormalization constants $Z(\beta_c(\theta_L)$ were obtained from dedicated $\theta=0$ simulations performed on hypercubic $N_s^4$ lattices (i.e., deep in the confined phase) for $\beta\simeq \beta_c$. All our results for $Z(\beta_c)$, $a(\beta_c) \sqrt(\sigma)$ and $T_c/\sqrt{\sigma}$, $\theta_I$ are reported in Tab.~\ref{tab:SU4_results}.

\begin{table*}[!htb]
\begin{tabular}{|c|c|c|c|c|c|c|c|}
\hline
\multicolumn{8}{|c|}{$N=4$}\\
\hline
&&&&&&&\\[-1em]
$N_s$ & $N_t$ & $\theta_L$ & $\theta_I$ & $\beta_c$ & $Z(\beta_c)$ & $a(\beta_c) \sqrt{\sigma}$ & $T_c / \sqrt{\sigma}$\\
\hline
\multirow{5}{*}{20} & \multirow{5}{*}{5}
&    0 & 0         & 10.634870(98) &  0.0850(13) &  0.32624(60) &  0.613048(77)(930)\\
& &  5 & 0.418(6)  & 10.637423(66) &  0.0837(12) &  0.32517(58) &  0.615065(52)(920)\\
& & 10 & 0.844(11) & 10.646041(97) &  0.0844(11) &  0.32160(55) &  0.621898(77)(880)\\
& & 15 & 1.303(19) & 10.661427(95) &  0.0869(13) &  0.31537(49) &  0.634185(77)(820)\\
& & 20 & 1.875(19) & 10.68176(11)  &  0.0938(14) &  0.30741(44) &  0.650594(89)(780)\\
\hline
\multirow{2}{*}{18} & \multirow{2}{*}{6}
  &  0 & 0         & 10.78625(42) &  0.1061(16) &  0.27106(48) &  0.61486(30)(110)\\
& & 20 & 2.377(34) & 10.85846(86) &  0.1189(17) &  0.24977(55) &  0.66728(63)(150)\\
\hline
\multirow{5}{*}{24} & \multirow{5}{*}{6}
  &  0 & 0         & 10.78676(15) &  0.1063(16) &  0.27090(48) &  0.61523(11)(110)\\
& &  5 & 0.543(8)  & 10.79062(97) &  0.1087(16) &  0.26969(49) &  0.61799(70)(110)\\
& & 10 & 1.113(15) & 10.8032(12)  &  0.1113(15) &  0.26581(50) &  0.62702(83)(120)\\
& & 15 & 1.700(26) & 10.82793(55) &  0.1134(17) &  0.25843(53) &  0.64492(40)(130)\\
& & 20 & 2.370(34) & 10.85749(94) &  0.1185(17) &  0.25004(55) &  0.66657(70)(150)\\
\hline
\multirow{5}{*}{24} & \multirow{5}{*}{8}
  &  0 & 0         & 11.0766(17)  &  0.1489(40) &  0.19978(55) &  0.62567(100)(230)\\
& &  5 & 0.724(19) & 11.08418(75) &  0.1448(38) &  0.19836(54) &  0.63018(44)(230)\\
& & 10 & 1.491(47) & 11.1113(14)  &  0.1491(47) &  0.19341(53) &  0.64630(84)(240)\\
& & 15 & 2.193(64) & 11.14530(61) &  0.1462(43) &  0.18751(51) &  0.66663(36)(240)\\
& & 20 & 3.161(98) & 11.2006(14)  &  0.1580(49) &  0.17859(50) &  0.69992(86)(260)\\
\hline
\end{tabular}
\caption{Summary of results obtained for $N=4$. The renormalization constants $Z(\beta_c(\theta_L))$ were obtained from dedicated $\theta=0$ simulations performed for $\beta\simeq \beta_c(\theta_L)$ on a $16^4$ lattice. The lattice spacings $a(\beta_c)\sqrt{\sigma}$ were obtained interpolating determinations of~\cite{Lucini:2005vg}. The critical temperatures are reported with two statistical errors: the first one is due to the uncertainty on $\beta_c$, the second one combines in quadrature the uncertainties on $\beta_c$ and $a(\beta_c)\sqrt{\sigma}$.}
\label{tab:SU4_results}
\end{table*}

\vspace{2\baselineskip}

\begin{figure}[!t]
\centering
\includegraphics[width=0.45\textwidth]{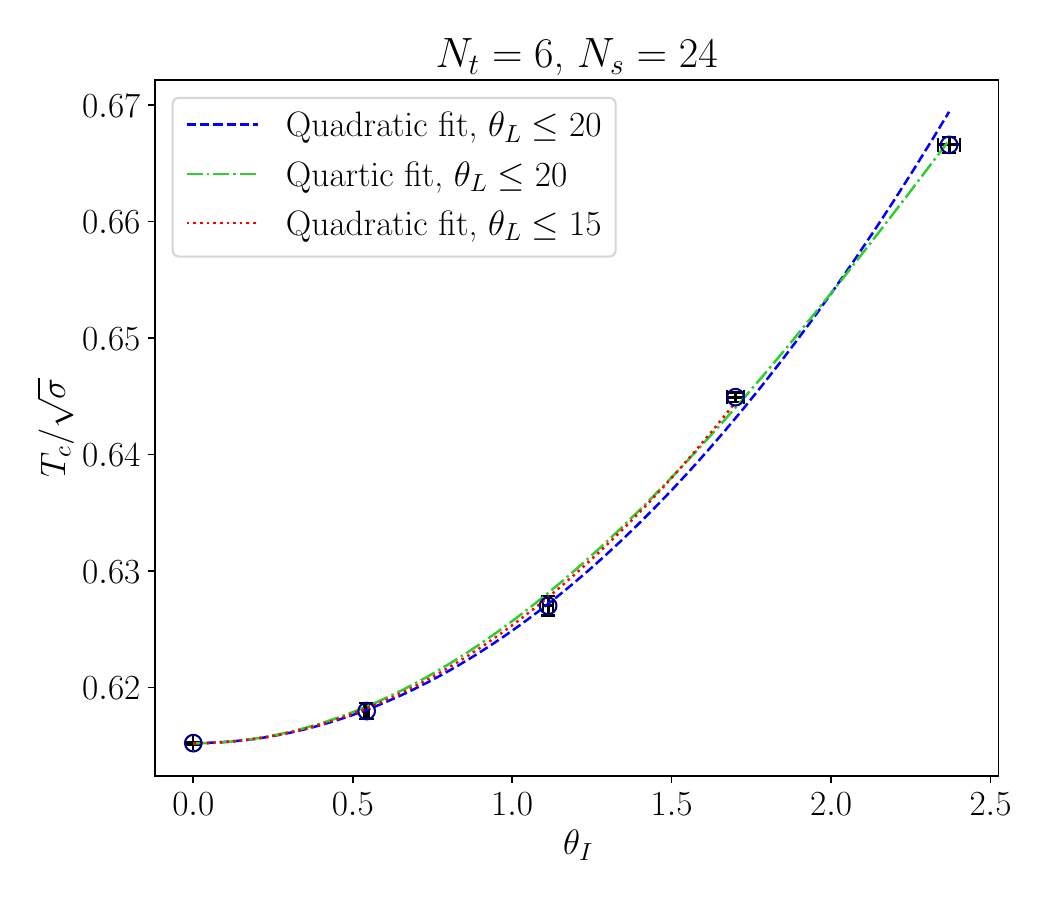}
\caption{Best fit of the imaginary-$\theta$ dependence of the critical temperature $T_c/\sqrt{\sigma}$ for $N=4$, determined on a $24^3 \times 8$ lattice, according to parabolic and quartic fit functions in $\theta_I$.}
\label{fig:SU4_fit_Tc_theta}
\end{figure}

We are now ready to compute the curvature $R$. Let us start from the computation of this quantity from the imaginary-$\theta$ dependence of the critical temperature. We fitted our data for $T_c/\sqrt{\sigma}$ as a function of $\theta_I$ according to the analytic continuation for imaginary values of $\theta$ of Eq.~\eqref{eq:Tc_thetadep_def}, which reads:
\beq\label{eq:fit_law}
\frac{T_c}{\sqrt{\sigma}}(\theta_I) = \frac{T_c}{\sqrt{\sigma}}[1+R\theta_I^2+O(\theta_I^4)],
\eeq
An example of the imaginary-$\theta$ fit of the critical temperature is shown in Fig.~\ref{fig:SU4_fit_Tc_theta}.

Concerning statistical errors on $R$, it is important to stress here that such best fit was carried out taking into account the errors on $\theta_I$, inherited from the ones on $Z(\beta_c)$, and the errors on $T_c/\sqrt{\sigma}$, inherited from the ones on $\beta_c$. As a matter of fact, although for what concerns the error on $T_c$ the contribution of the uncertainty on $\beta_c$ is largely sub-dominant with respect to the error on the scale setting, we found the opposite to hold for $R$, being the values of $a(\beta_c)\sqrt{\sigma}$ highly-correlated among themselves (since they all come from an interpolation of the data of~\cite{Lucini:2005vg} for the string tension). In particular, a bootstrap analysis done to take into account correlations among $a(\beta_c)\sqrt{\sigma}$ reveals that the joint fluctuations of $a(\beta_c)\sqrt{\sigma}$ just have the effect of overall shifting the data for $T_c(\theta_I)$, with a negligible impact on their curvature, eventually leading to a small uncertainty on $R$.

\begin{figure}[!t]
\centering
\includegraphics[width=0.48\textwidth]{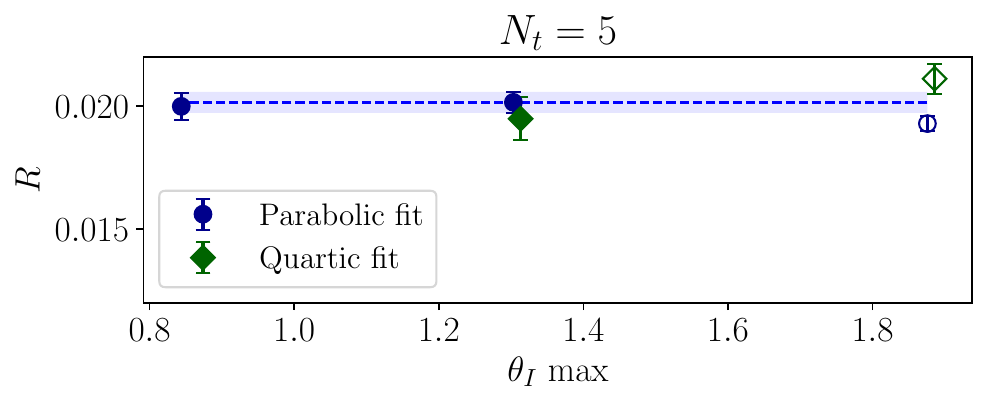}
\includegraphics[width=0.48\textwidth]{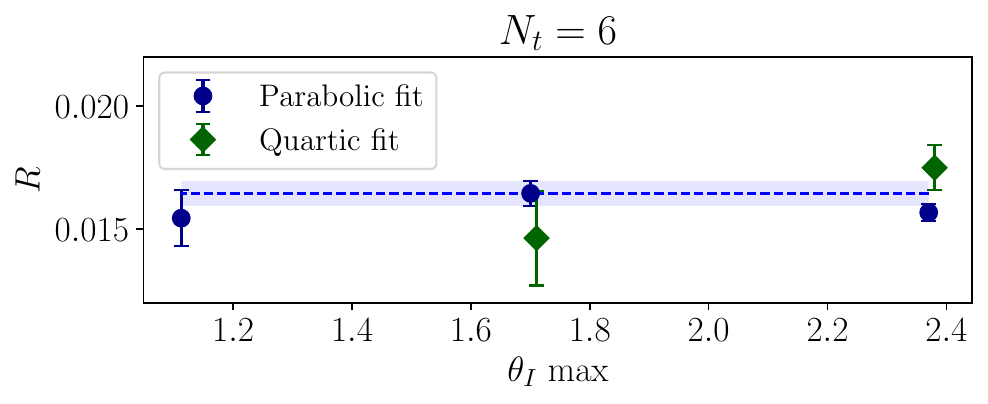}
\includegraphics[width=0.48\textwidth]{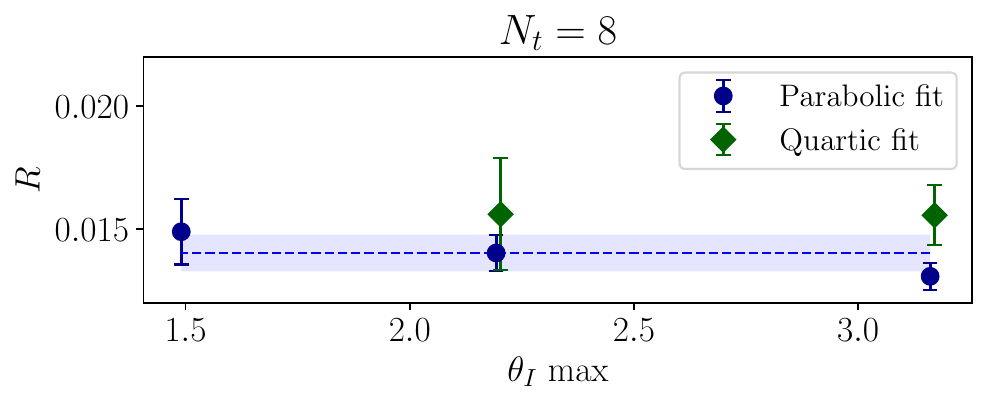}
\caption{Study of the systematic effects on $R(N=4)$ due to the truncation of the Taylor expansion in $\theta_I$. The solid lines and the shaded areas represent our final results. Empty points were ignored, as they are associated with best fits with a $p$-value smaller than 5\%.}
\label{fig:SU4_curv_systematics}
\end{figure}

Concerning instead systematic errors, since we are dealing with a truncated Taylor expansion, one needs to check that systematic effects coming from the neglected terms are under control. Thus, to check that contributions from higher-order terms are indeed small, we performed both parabolic and quartic fits in $\theta_I$ for various choices of the fit range, cf.~Fig.~\ref{fig:SU4_curv_systematics}. In all cases, systematics due to higher-order powers of $\theta_I$ are well under control within our typical statistical errors, see Fig.~\ref{fig:SU4_curv_systematics} and Tab.~\ref{tab:SU4_curv}. Our final results for $R$ from the imaginary-$\theta$ fits of the critical temperature are reported for each $N_t$ in Tab.~\ref{tab:FINAL_RES_SU4}.

\begin{table}[!htb]
\begin{tabular}{|c|c|c|c|c|c|c|}
\hline
\multicolumn{7}{|c|}{$N=4$}\\
\hline
&&&&&&\\[-1em]
$N_s$ & $N_t$ & $\theta_{I,\,\max}$ & $R$ & $\tilde{\chi}^2$ & $\mathrm{dof}$ & $p$-value\\
\hline
\multicolumn{7}{|c|}{Quadratic fit}\\
\hline
\multirow{3}{*}{20} & \multirow{3}{*}{5}
    &  0.84 &  0.01999(56) &  1.80 & 1 &  18.02\% \\
 &  &  1.30 &  0.02015(43) &  2.00 & 2 &  36.81\% \\
 &  &  1.88 &  0.01929(32) * &  19.94 & 3 &  0.02\% \\
\hline
\multirow{3}{*}{24} & \multirow{3}{*}{6}
    &  1.11 &  0.0155(11) &  0.01 & 1 &  93.85\% \\
 &  &  1.70 &  0.01645(51) &  0.96 & 2 &  61.80\% \\
 &  &  2.37 &  0.01567(35) &  6.44 & 3 &  9.21\% \\
\hline
\multirow{3}{*}{24} & \multirow{3}{*}{8}
    &  1.49 &  0.0149(13)  &  0.13 & 1 &  71.85\% \\
 &  &  2.19 &  0.01403(75) &  0.90 & 2 &  63.74\% \\
 &  &  3.16 &  0.01308(54) &  5.45 & 3 &  14.17\% \\
\hline
\multicolumn{7}{|c|}{Quartic fit}\\
\hline
\multirow{2}{*}{20} & \multirow{2}{*}{5}
    &  1.30 &  0.01949(87) &  1.24 & 1 &  26.55\% \\
 &  &  1.88 &  0.02111(60) * &  7.40 & 2 &  2.47\% \\
\hline
\multirow{2}{*}{24} & \multirow{2}{*}{6}
    &  1.70 &  0.0146(19)  &  0.01 & 1 &  92.97\% \\
 &  &  2.37 &  0.01750(91) &  2.86 & 2 &  23.94\% \\
\hline
\multirow{2}{*}{24} & \multirow{2}{*}{8}
    &  2.19 &  0.0156(23) &  0.35 & 1 &  55.22\% \\
 &  &  3.16 &  0.0156(12) &  0.36 & 2 &  83.70\% \\
\hline
\end{tabular}
\caption{Results of the imaginary-$\theta$ fit of $T_c/\sqrt{\sigma}$ as a function of $\theta_I$ for $N=4$ and for various fit ranges, and considering both a parabolic and a quartic fit function in $\theta_I$. Results marked with * were not considered when assessing the systematic error on $R$ because of the unreliability of the fits.}
\label{tab:SU4_curv}
\end{table}

\begin{table}[!t]
\begin{tabular}{|c|c|c|c|c|c|c|}
\hline
\multicolumn{7}{|c|}{$N=4$}\\
\hline
\multicolumn{5}{|c|}{Latent heat $\theta=0$ runs} & & \multicolumn{1}{c|}{Imaginary-$\theta$ runs}\\
\cline{1-5}\cline{7-7}
&&&&&&\\[-1em]
$N_s$ & $N_t$ & $\dfrac{\Delta \chi}{T_c^4}$ & $\dfrac{\Delta \epsilon}{T_c^4}$ & $R=\dfrac{\Delta\chi}{2\Delta\epsilon}$ && $R$ (imaginary-$\theta$ fit) \\
&&&&&&\\[-1em]
\cline{1-5}\cline{7-7}
&&&&&&\\[-1em]
20 & 5 & 0.1598(34) & 5.357(92) & 0.01489(36) &&  0.0202(4)\\
36 & 6 & 0.1211(57) & 4.373(55) & 0.01385(67) &&  0.0165(5)\\
48 & 8 & 0.117(13)  & 4.466(55) & 0.0131(14)  &&  0.0140(8)\\
\hline
\end{tabular}
\caption{Summary of our final results for the curvature $R$ for $N=4$ obtained from dedicated $\theta=0$ simulations aiming at determining the jump of the topological susceptibility $\Delta \chi$ and the latent heat $\Delta \epsilon$. For completeness we also report the final results for $R$ obtained from the imaginary-$\theta$ fit.}
\label{tab:FINAL_RES_SU4}
\end{table}

We now move to the determinations of $R$ from $\theta_L=0$ simulations alone, using Eq.~\eqref{eq:R_conjecture}. In order to reliably compute the jump of the topological susceptibility according to Eq.~\eqref{eq:jump_chi} and the latent heat according to Eq.~\eqref{eq:latent_heat}, which require to compute differences of expectation values between the confined and the deconfined phases, we performed (for all values of $N_t$ but the smaller) dedicated $\theta=0$ simulations at $\beta \simeq \beta_c$ on lattices with a larger spatial volume compared to those employed for the identification of the critical point, the purpose being of obtaining a sharper separation between the two phases.

\begin{figure}[!t]
\centering
\includegraphics[width=0.4\textwidth]{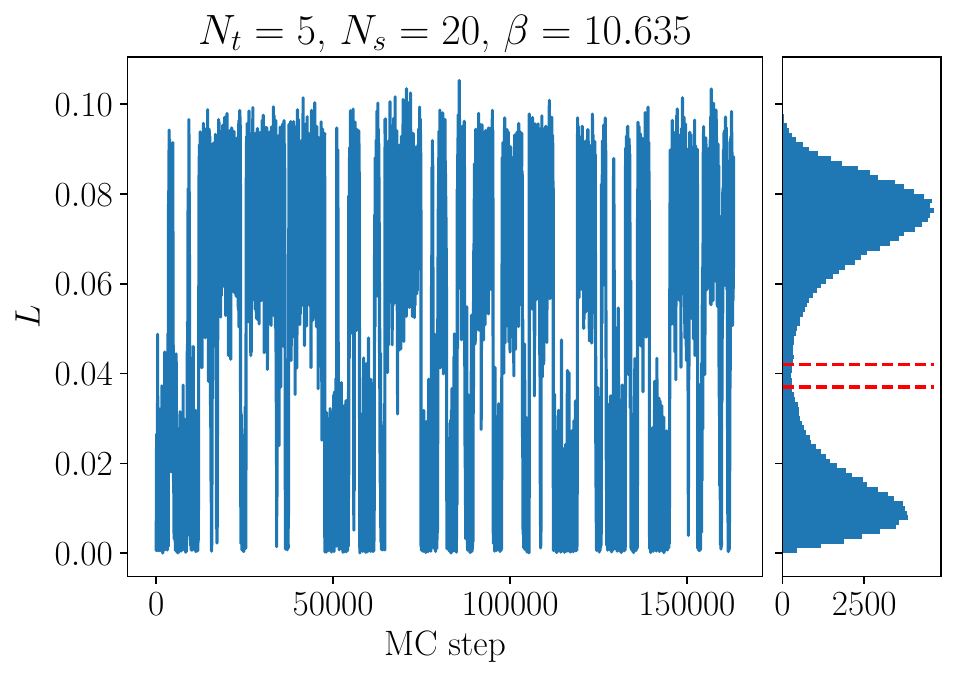}
\includegraphics[width=0.4\textwidth]{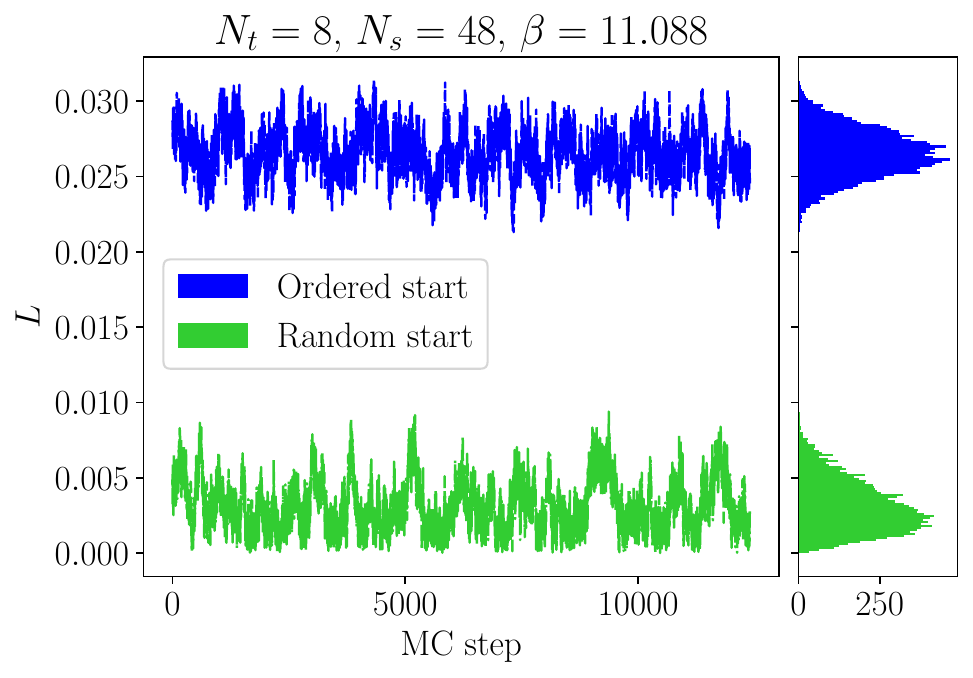}
\caption{Top panel: history and histogram of the Polyakov loop $L$ for $N=4$ obtained on a $20^3 \times 5$ for $\beta\simeq \beta_c(\theta=0)$. The two solid lines correspond to the cuts $L<L_{\mathrm{cut}}^{(c)}$ and $L>L_{\mathrm{cut}}^{(d)}$ used to assign a configuration to the confined or deconfined phase. Bottom panel: histories of the Polyakov loop $L$ for $N=4$ obtained on a $48^3 \times 8$ for $\beta\simeq \beta_c(\theta=0)$ and starting respectively from an ordered/random configuration. In both cases, the horizontal MC time is expressed in terms of the number of single MC steps defined at the beginning of Sec.~\ref{sec:algos}.}
\label{fig:poly_histo_2peaks_large_vol_N4}
\end{figure}

An example of the histories of the Polyakov loop obtained from such simulations on a $48 \times 8$ lattice is shown in the bottom panel of Fig.~\ref{fig:poly_histo_2peaks_large_vol_N4}. As it can be observed, by starting from ordered/random configurations we obtain histories which explore only a single phase, thus allowing an unambiguous calculation of the expectation values $\braket{\mathcal{O}}_{c,d}$ in each of the two phases. 

On the other hand, for the coarsest lattice spacing explored (corresponding to $N_t=5$) the latent heat turned out to be large enough to allow an unambiguous separation of the two phases already on the volume employed to compute $\beta_c$, as it can be observed by inspecting the histogram of the Polyakov loop in that case, shown in the top panel of Fig.~\ref{fig:poly_histo_2peaks_large_vol_N4}. Therefore, in that case we just fixed two cuts, $L_{\mathrm{cut}}^{(c)}$ and $L_{\mathrm{cut}}^{(d)}$, and assigned configurations with $L<L_{\mathrm{cut}}^{(c)}$ and $L>L_{\mathrm{cut}}^{(d)}$, to, respectively, the confined and the deconfined phase. We also verified that our results for $\Delta \epsilon(N_t=5)$ were largely insensitive to the specific choice of these cuts as long as their values were varied within the depleted region separating the two peaks, cf.~top panel of Fig.~\ref{fig:poly_histo_2peaks_large_vol_N4}.

Our results for the jump of the topological susceptibility $\Delta \chi/T_c^4$, the latent heat $\Delta \epsilon/T_c^4$ and the curvature $R=\Delta\chi/(2 \Delta \epsilon)$ according to Eq.~\eqref{eq:R_conjecture} for all explored points are reported in Tab.~\ref{tab:FINAL_RES_SU4}, along with the final results for $R$ obtained from the imaginary-$\theta$ fit of $T_c$. We now proceed to perform the extrapolation of both determinations of $R$ towards the continuum limit. In both cases, we took the continuum limit assuming $O(a^2)$ corrections, i.e., since $1/N_t = a T_c$, according to the law:
\beq\label{eq:fit_continuum}
R(N_t) = R + c/N_t^2,
\eeq
where $R$ represents the continuum-extrapolated result.

\begin{figure}[!t]
\centering
\includegraphics[width=0.4\textwidth]{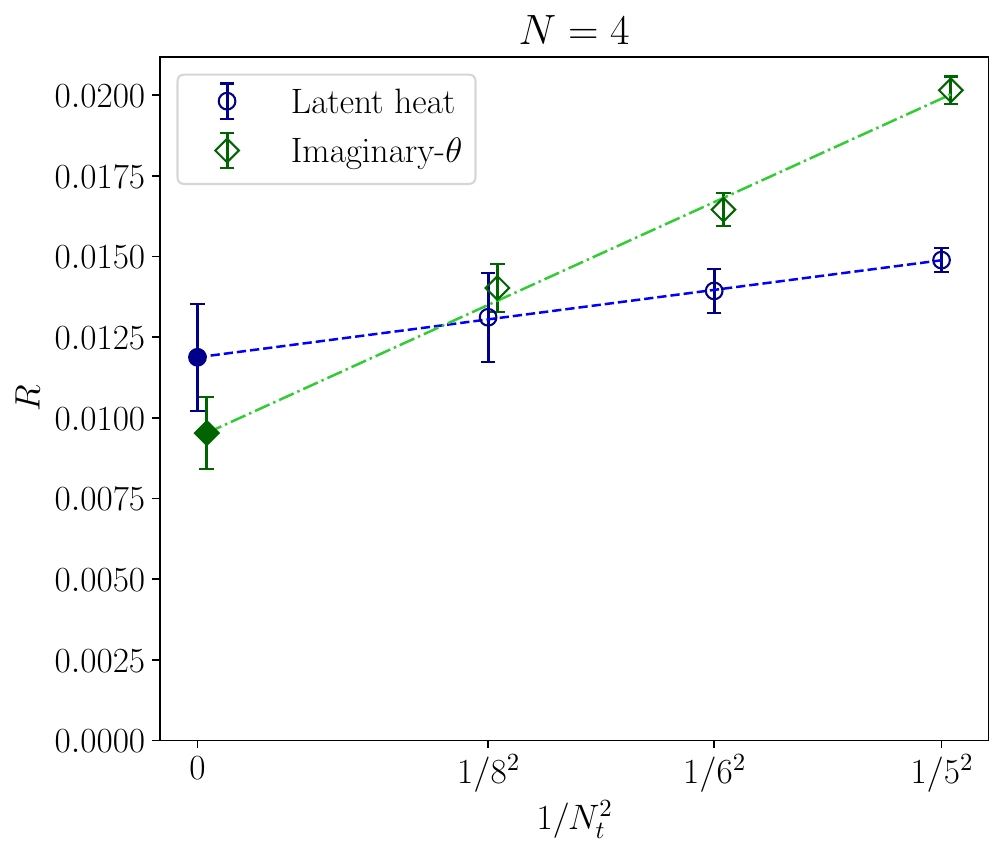}
\includegraphics[width=0.4\textwidth]{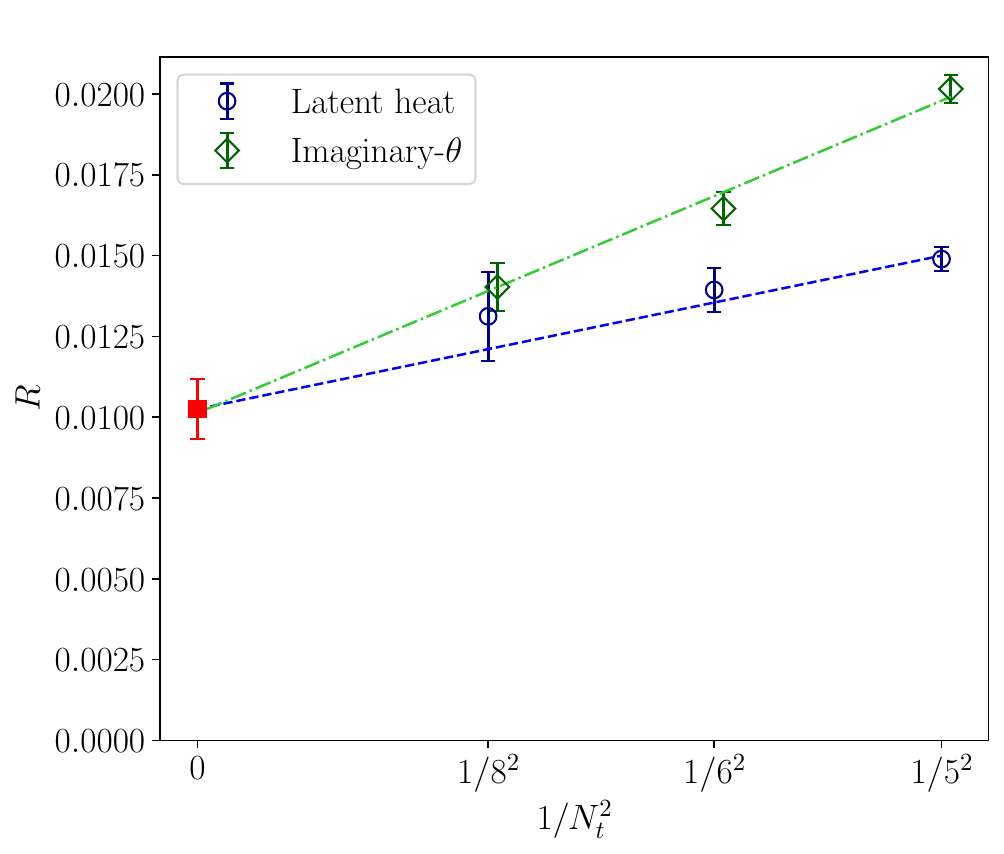}
\caption{Extrapolation towards the continuum limit according to~\eqref{eq:fit_continuum} of the determinations of the curvature $R(N=4)$ in Tab.~\ref{tab:FINAL_RES_SU4} by, respectively, fitting the two data sets separately (top), and imposing a common continuum limit (bottom).}
\label{fig:cont_limit_N4}
\end{figure}

The continuum limits of $R$ from both data sets are shown and compared in the top panel of Fig.~\ref{fig:cont_limit_N4}. Our final continuum results are:
\beq
R(N=4) &= 0.0095(11) \quad &\text{ (from imaginary-$\theta$),}\\
\nonumber\\
R(N=4) &= 0.0119(16) \quad &\text{ (from latent heat)}.
\eeq
As it can be observed, obtained determinations are perfectly compatible within the errors once the continuum limit is taken, confirming that, also for $N>3$, the predicted relation between the curvature of the $T-\theta$ phase diagram and the jump at the critical point of the internal energy and the topological susceptibility holds.

Given the very good agreement between these two results, we also tried to take a common continuum limit, where a common fit parameter was used for $R$ in Eq.~\eqref{eq:fit_continuum}. The two data set can be nicely fitted jointly with such a fit function, and the best fit yields a reduced chi-squared of $2.3/3$, corresponding to a $p$-value of $52\%$, and a combined continuum limit:
\beq\label{eq:cont_lim_combo_N4}
R(N=4) = 0.01025(92) \quad \text{ (combined)}.
\eeq
The combined fit is shown in the bottom panel of Fig.~\ref{fig:cont_limit_N4}.

\subsection{Results for $\SU(6)$}

The computation of $R$ for $N=6$ goes essentially along the same lines outlined for $\SU(4)$, meaning that we looked for the critical couplings $\beta_c(\theta_L)$ for several values of the imaginary-$\theta$ parameter, and for 3 values of the temporal extent $N_t=5,6,8$. However, compared to the $N=4$, there are two important differences that we want to stress before presenting our numerical results.

The first difference relies on the adopted volumes, which are smaller than those chosen for $N=4$. As a matter of fact, to determine $\beta_c(\theta_L)$ for $N=6$ we adopted in all cases lattices with aspect ratios $N_s/N_t=2$. Indeed, being $\Delta \epsilon \sim N^2$, the strength of the first order for $N=6$ was such that, on lattices with larger volumes, we were not able to unambiguously identify the critical point. Nonetheless, on the basis of general theoretical arguments relying on large-$N$ volume reduction~\cite{PhysRevLett.48.1063}, we expect finite size effects to become milder at large $N$. This fact has also been verified in previous lattice studies of the $\SU(N)$ phase transition~\cite{Lucini:2005vg,Lucini:2012wq}. Moreover, the curvature $R$ is essentially the leading-order deviation from 1 of $T_c(\theta)/T_c(0)$, thus, finite size effects for this quantity are expected to largely drop in the ratio. In conclusion, we are confident that, even though smaller spatial volumes were used compared to the $N=4$ case, our determinations of $R$ in the $N=6$ case are still reliable.

The second difference with respect to the $N=4$ case, is that, as already discussed in Sec.~\ref{sec:algos}, for $N=6$ the simulations performed for larger values of $\beta$ suffer from a severe topological freezing. In particular, we found this problem to affect all simulations performed on the $16^3 \times 8$ lattice, for all values of $\beta$ and $\theta_L$ employed. Thus, all simulations performed on this lattice for $N=6$ were done using the PTBC algorithm, which provides a tremendous gain in terms of the autocorrelation time of the topological charge. To be more precise, for each value of $\beta$ and $\theta_L$ explored at $N_t=8$ we simulated $N_r = 30$ replicas of the lattice, and boundary conditions were changed on a cubic defect of size $L_d = 3$. The tempering parameters used to change boundary conditions were tuned through short test runs to ensure an approximately constant replica swap rate~\eqref{eq:PTBC_swap_rate} of $\sim 30\%$.

\begin{figure}
\centering
\includegraphics[width=0.48\textwidth]{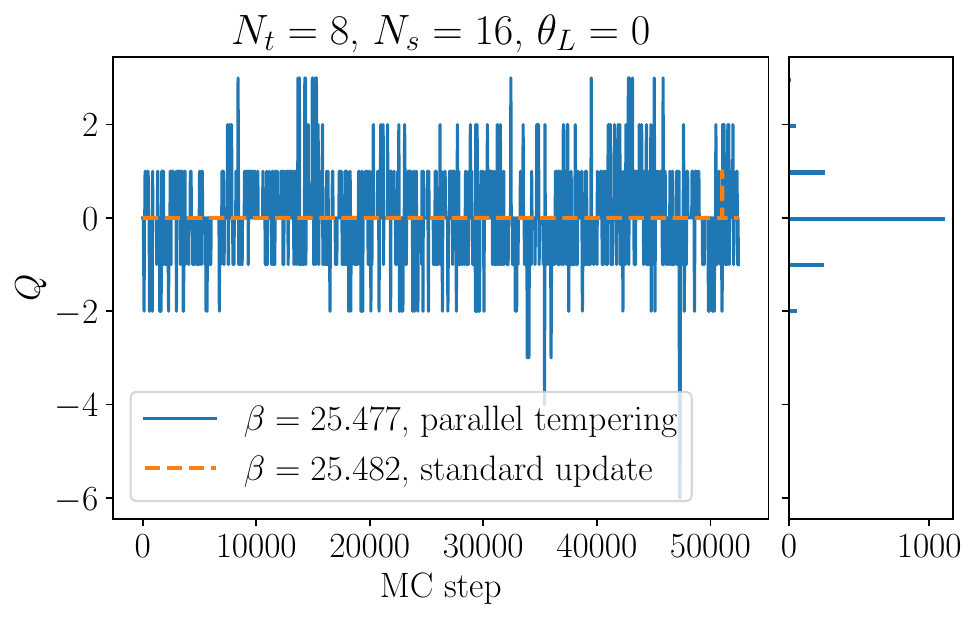}
\includegraphics[width=0.48\textwidth]{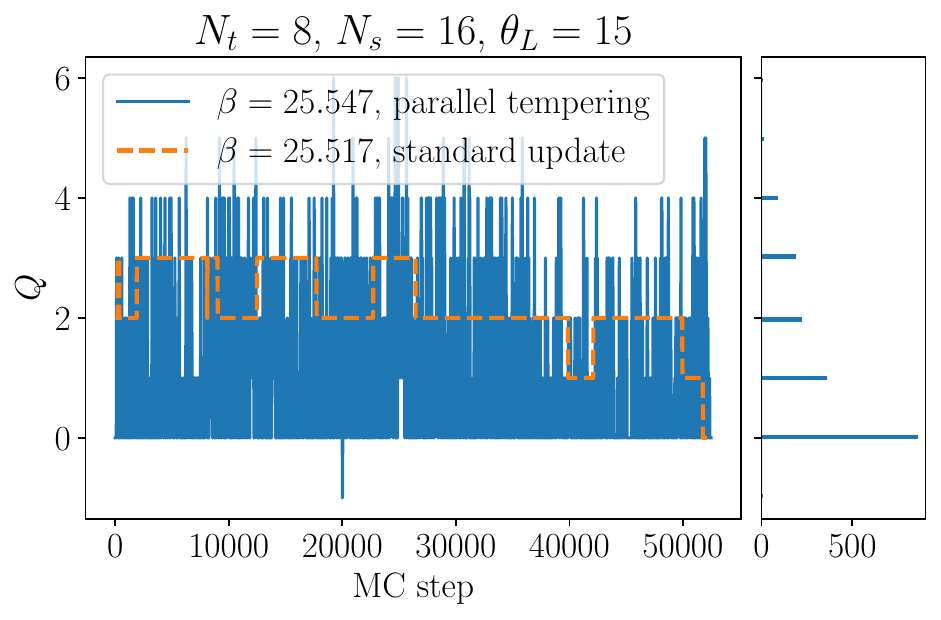}
\caption{Comparison of the evolutions of the topological charge $Q$ defined in Eq.~\eqref{eq:rounded_topcharge} obtained with PTBC and with the standard algorithm for the largest $\beta$ explored at $\theta_L=0$ and $\theta_L=15$ on a $16^3 \times 8 $ lattice for $N=6$. The horizontal MC time is expressed in both cases in terms of the number of stadard MC steps defined at the beginning of Sec.~\ref{sec:algos}. For the PTBC algorithm this means, in practice, to multiply the number of parallel tempering steps by the number of replicas.}
\label{fig:topcharge_evo_N6}
\end{figure}

\begin{figure}
\centering
\includegraphics[width=0.48\textwidth]{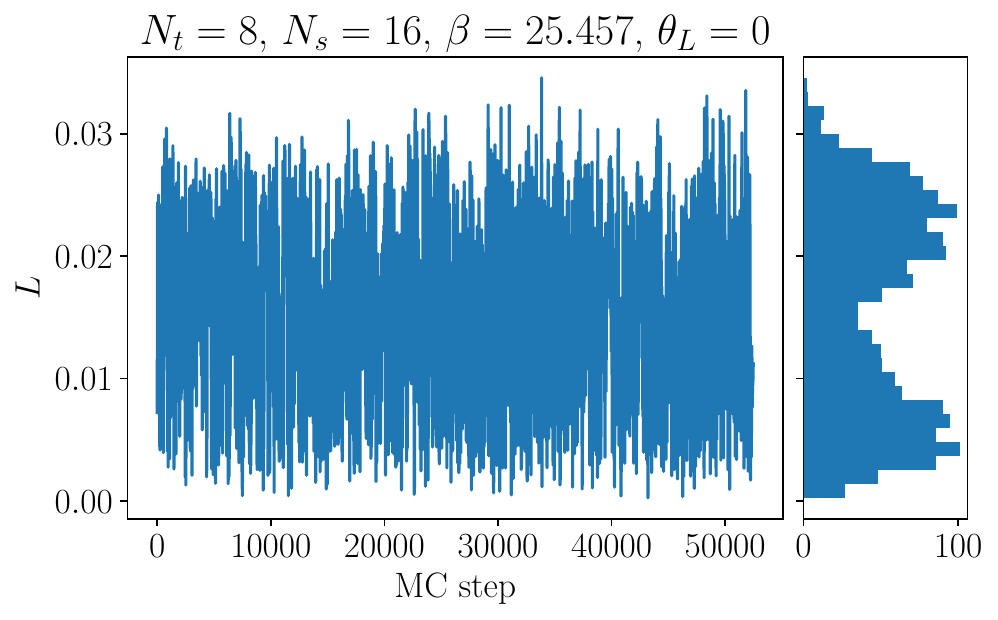}
\includegraphics[width=0.48\textwidth]{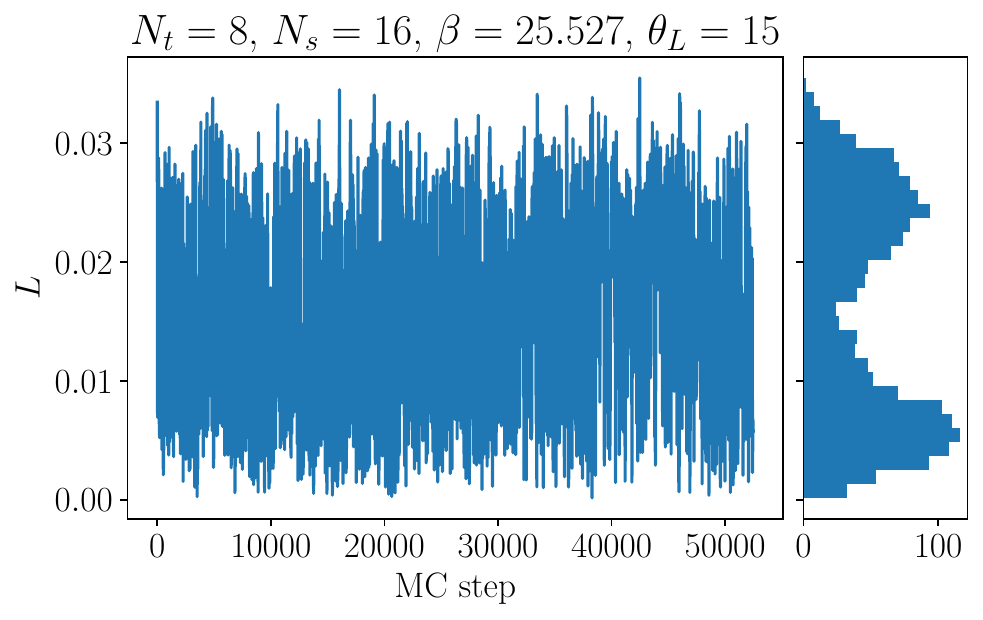}
\caption{Evolution of the Polyakov loop $L$ obtained with parallel tempering for $\beta \simeq \beta_c$ and $\theta_L=0, 15$ on a $16^3 \times 8$ lattice for $N=6$. The horizontal MC time is expressed in terms of the number of standard MC steps defined at the beginning of Sec.~\ref{sec:algos}, i.e., the number of parallel tempering steps have been multiplied by the number of replicas.}
\label{fig:poly_evo_N6}
\end{figure}

Let us start exactly by the latter point. In Fig.~\ref{fig:topcharge_evo_N6}, we show the history of the topological charge obtained from the standard combination of local algorithms, and from the parallel tempering. As it can be observed, the PTBC algorithm allows to obtain a substantial gain in terms of the observed number of fluctuations of $Q$ with the same numerical effort\footnote{The history at non-zero $\theta_L$ shown in Fig.~\ref{fig:topcharge_evo_N6} presents a remarkable feature: negative values of $Q$ are largely suppressed. This is a feature common also to other MC histories obtained for non-zero $\theta_L$ and
smaller values of $N_t$, i.e., without the PTBC algorithm, which can be easily interpreted as a large-$N$, finite-$T$ and finite-$\theta_L$ phenomenon as follows. 
The explored temperature, in absence of $\theta_L$, would be in the deconfined phase, where fluctuations of $Q$ are exponentially suppressed with $N$. 
The addition of a non-zero $\theta_L > 0$ enhances the probability of exploring non-zero values of $Q$, thus trying to bring the system back into the confined phase, but only non-zero values of $Q$ with the correct sign are enhanced. Therefore, fluctuations of $Q$ with the wrong sign are doubly suppressed, both by the large value of $N$ and by the imaginary $\theta$-parameter.}. Remarkably enough, we also found that, with the PTBC algorithm, it is possible to achieve a reduction of about a factor of $3$ of the autocorrelation time of $L$. This is probably due to the large correlation between $L$ and $Q$. Examples of MC histories of the Polyakov loop are shown in Fig.~\ref{fig:poly_evo_N6}, where for the sake of clarity we just report the MC evolution of $L$ obtained with parallel tempering.

\begin{figure*}[!htb]
\centering
\includegraphics[width=0.9\textwidth]{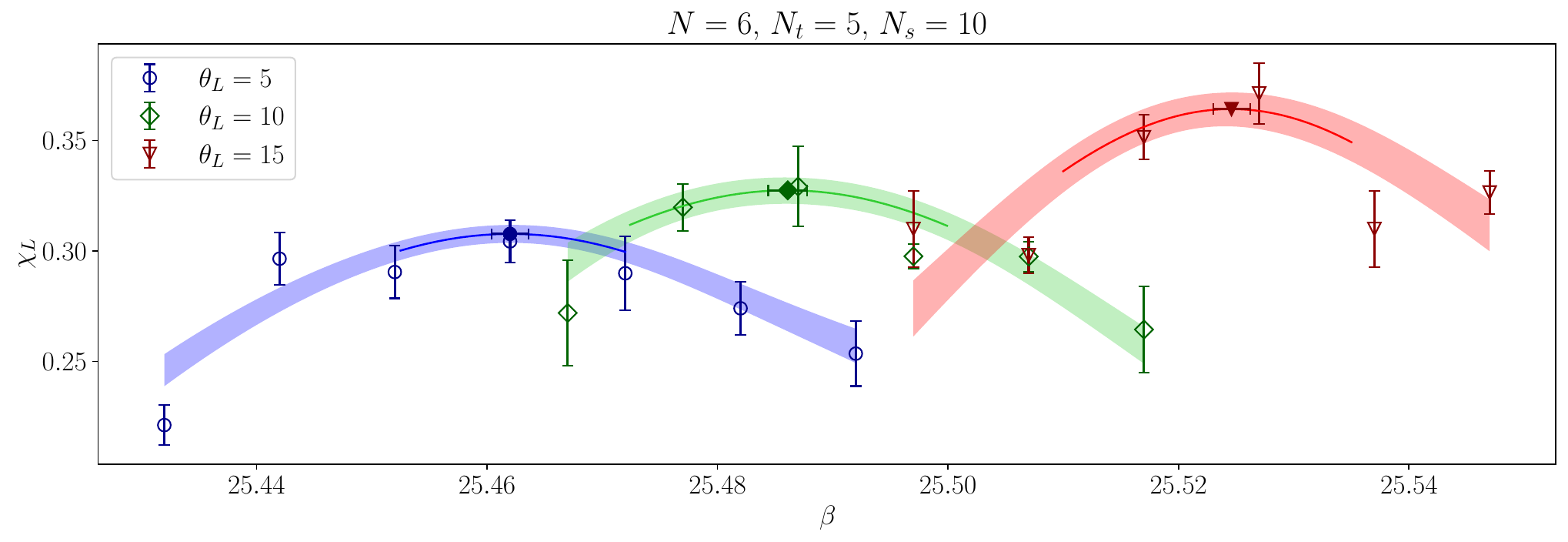}
\caption{Peaks of the Polyakov loop susceptibility $\chi_L(\beta)$ as a function of the bare coupling $\beta$ observed on a $12^3\times6$ lattice for 3 different values of $\theta_L=0,15,20$ and $N=6$. The lighter shaded areas represent the multihstogram interpolations of the MC data, while the darker solid lines are the results of the best fits of the interpolated data according to a Lorentzian ansatz $f(\beta)=A_1/[1+(\beta-\beta_c)^2/A_2]$. Filled points represent the position of the critical points. }
\label{fig:chi_L_peaks_N6}
\end{figure*}

We now move to the determination of $R$ from the imaginary-$\theta$ fit. In Fig.~\ref{fig:chi_L_peaks_N6} we show how the peaks of $\chi_L(\beta)$ shift towards larger couplings as $\theta_L$ is increased. Our determinations of $\beta_c(\theta_L)$ for $N=6$ are collected in Tab.~\ref{tab:SU6_results}, along with the corresponding values of $T_c/\sqrt{\sigma}$.

\begin{figure}[!t]
\centering
\includegraphics[width=0.48\textwidth]{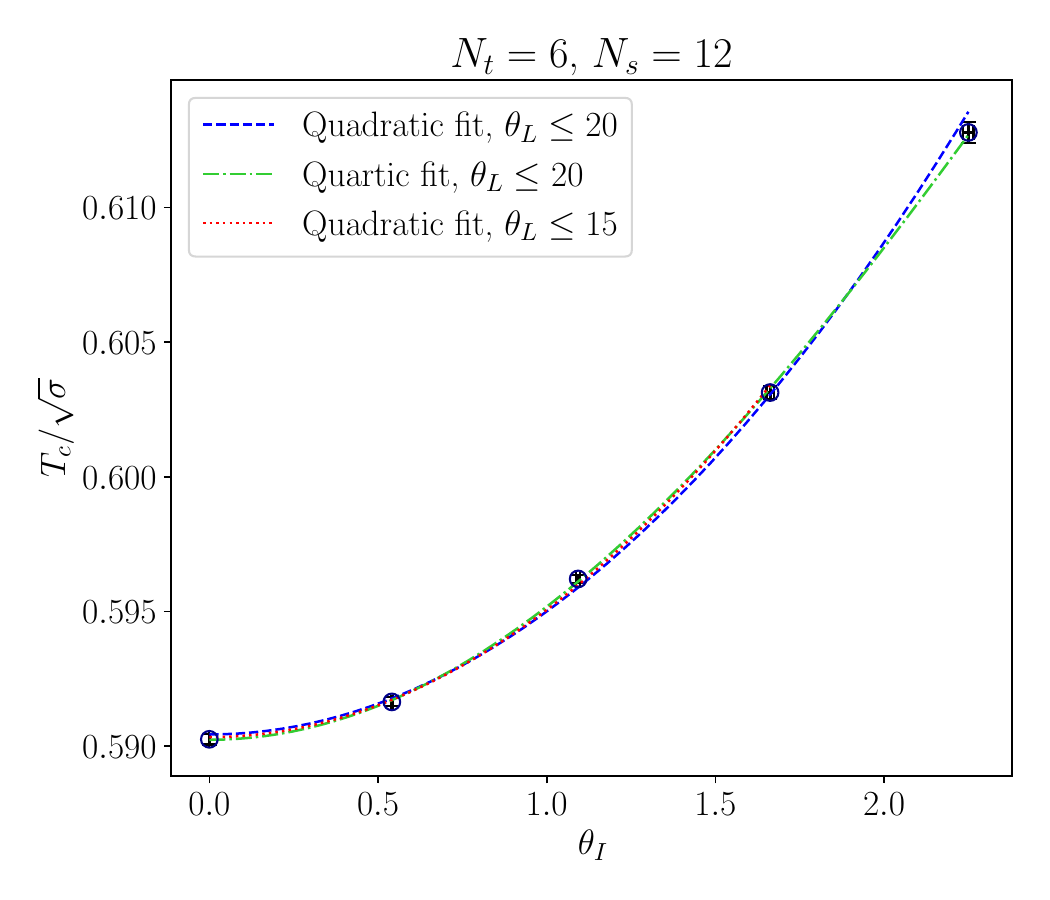}
\caption{Best fit of the imaginary-$\theta$ dependence of the critical temperature $T_c/\sqrt{\sigma}$ for $N=6$, determined on a $12^3 \times 6$ lattice, according to parabolic and quartic fit functions in $\theta_I$.}
\label{fig:SU6_fit_Tc_theta}
\end{figure}

A few examples of the imaginary-$\theta$ fit of $T_c/\sqrt{\sigma}$ for $N=6$ results are shown instead in Fig.~\ref{fig:SU6_fit_Tc_theta}. Again, in order to assess possible systematic effects on the curvature due to the truncation of the Taylor series, we performed both parabolic and quadratic fits in $\theta_I$ for various choices of the fit range. All obtained results are collected in Tab.~\ref{tab:SU6_comp}.

\begin{table*}[!t]
\begin{tabular}{|c|c|c|c|c|c|c|c|}
\hline
\multicolumn{8}{|c|}{$N=6$}\\
\hline
&&&&&&&\\[-1em]
$N_s$ & $N_t$ & $\theta_L$ & $\theta_I$ & $\beta_c$ & $Z(\beta_c)$ & $a(\beta_c) \sqrt{\sigma}$ & $T_c / \sqrt{\sigma}$ \\
\hline
\multirow{5}{*}{10} & \multirow{5}{*}{5}
  &  0 & 0          & 24.49630(16) &  0.08306(24) &  0.3423(11) &  0.5842(1)(16)\\
& &  5 & 0.4161(12) & 24.49827(32) &  0.08322(24) &  0.3419(11) &  0.5849(1)(16)\\
& & 07 & 0.5842(18) & 24.50089(33) &  0.08346(25) &  0.3414(11) &  0.5859(1)(16)\\
& & 10 & 0.8403(27) & 24.50712(30) &  0.08404(27) &  0.3400(11) &  0.5881(1)(16)\\
& & 15 & 1.2785(47) & 24.52017(31) &  0.08523(31) &  0.3373(10) &  0.5930(1)(15)\\
\hline
\multirow{5}{*}{12} & \multirow{5}{*}{6}
  &  0 & 0         & 24.82491(59) &  0.10785(63) &  0.28236(76) &  0.5902(2)(16)\\
& &  5 & 0.541(3)  & 24.82922(51) &  0.10821(64) &  0.28170(76) &  0.5916(2)(16)\\
& & 10 & 1.093(7)  & 24.84340(47) &  0.10933(68) &  0.27954(77) &  0.5962(2)(16)\\
& & 15 & 1.662(11) & 24.86484(71) &  0.11082(72) &  0.27633(77) &  0.6031(2)(17)\\
& & 20 & 2.250(15) & 24.8947(12)  &  0.11251(76) &  0.27198(78) &  0.6128(4)(18)\\
\hline
\multirow{4}{*}{16} & \multirow{4}{*}{8}
&  0 & 0            & 25.45617(92)&  0.1414(12) &  0.20956(83) &  0.5965(2)(31)\\
& &  5 & 0.7086(64) & 25.4620(16) &  0.1417(13) &  0.20903(83) &  0.5979(3)(32)\\
& &  7 & 0.9953(90) & 25.4725(21) &  0.1422(13) &  0.20817(86) &  0.6005(5)(33)\\
& & 10 & 1.428(13)  & 25.4861(17) &  0.1428(13) &  0.20704(89) &  0.6037(4)(34)\\
& & 15 & 2.166(20)  & 25.5246(16) &  0.1444(13) &  0.20394(99) &  0.6129(4)(40)\\
\hline
\end{tabular}
\caption{Summary of results obtained for $N=6$. The renormalization constant $Z(\beta_c(\theta_L))$ were obtained from interpolating the results for $Z(\beta)$ reported in Refs.~\cite{Bonati:2016tvi,Bonanno:2020hht}. The lattice spacings $a(\beta_c)\sqrt{\sigma}$ were obtained interpolating determinations of~\cite{Lucini:2005vg}. The critical temperatures are reported with two statistical errors: the first one is due to the uncertainty on $\beta_c$, the second one combines in quadrature the uncertainties on $\beta_c$ and $a(\beta_c)\sqrt{\sigma}$.}
\label{tab:SU6_results}
\end{table*}

\begin{figure}[!htbp]
\centering
\includegraphics[width=0.48\textwidth]{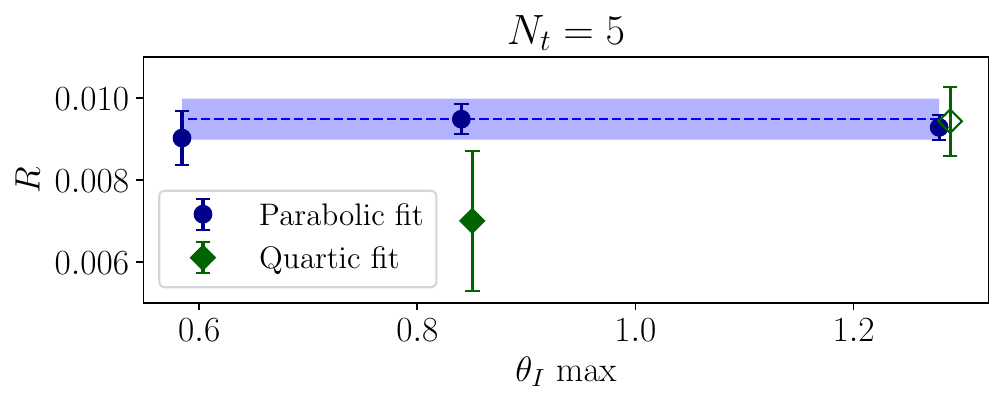}
\includegraphics[width=0.48\textwidth]{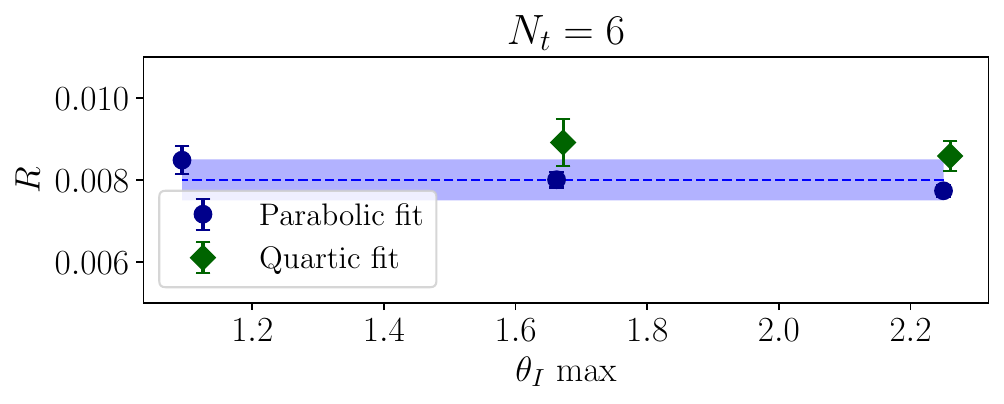}
\includegraphics[width=0.48\textwidth]{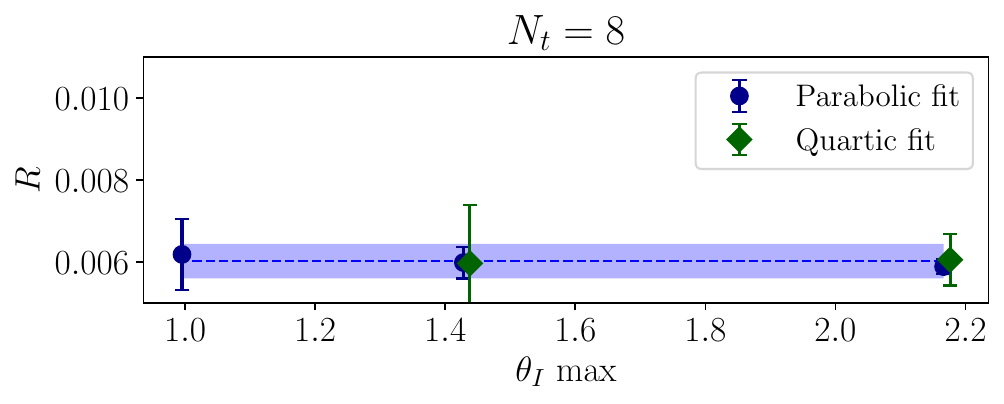}
\caption{Study of the systematic effects on $R(N=6)$ due to the truncation of the Taylor expansion in $\theta_I$. The solid lines and the shaded are represent our final results. Empty points were ignored, as they are associated with best fits with a $p$-value smaller than 5\%.}
\label{fig:SU6_curv_systematics}
\end{figure}

\begin{table}[!htb]
\begin{tabular}{|c|c|c|c|c|c|c|}
\hline
\multicolumn{7}{|c|}{$N=6$}\\
\hline
&&&&&&\\[-1em]
$N_s$ & $N_t$ & $\theta_{I \, max}$ & $R$ & $\chi^2$ & n dof & $p_{val}$\\
\hline
\multicolumn{7}{|c|}{Quadratic fit}\\
\hline
\multirow{3}{*}{10} & \multirow{3}{*}{5}
   &  0.58 &  0.00902(66)  &  0.89 & 1 &  34.42\% \\
&  &  0.84 &  0.00948(37) &  5.88 & 2 &   5.28\% \\
&  &  1.28 &  0.00928(30) &  6.53 & 3 &   8.84\% \\
\hline
\multirow{3}{*}{12} & \multirow{3}{*}{6}
   &  1.09 &  0.00848(34) &  0.10 & 1 &  75.78\% \\
&  &  1.66 &  0.00800(19) &  3.08 & 2 &  21.44\% \\
&  &  2.25 &  0.00773(14) &  7.52 & 3 &   5.71\% \\
\hline
\multirow{3}{*}{16} & \multirow{3}{*}{8}
   &  1.00 &  0.00618(86) &  0.73 & 1 &  39.23\% \\
&  &  1.43 &  0.00598(39) &  1.04 & 2 &  59.46\% \\
&  &  2.17 &  0.00588(18) &  1.17 & 3 &  75.93\% \\
\hline
\multicolumn{7}{|c|}{Quartic fit}\\
\hline
\multirow{2}{*}{10} & \multirow{2}{*}{5}
   &  0.84 &  0.0070(17)  &  0.24 & 1 &  62.51\% \\
&  &  1.28 &  0.00943(84) * &  6.42 & 2 &   4.04\% \\
\hline
\multirow{2}{*}{12} & \multirow{2}{*}{6}
   &  1.66 &  0.00891(58) &  0.31 & 1 &  57.71\% \\
&  &  2.25 &  0.00858(36) &  0.84 & 2 &  65.84\% \\
\hline
\multirow{2}{*}{16} & \multirow{2}{*}{8}
   &  1.43 &  0.0059(14)  &  0.90 & 1 &  34.39\% \\
&  &  2.17 &  0.00605(63) &  0.99 & 2 &  60.95\% \\
\hline
\end{tabular}
\caption{Results of the imaginary-$\theta$ fit of $T_c/\sqrt{\sigma}$ as a function of $\theta_I$ for $N=6$ and for various fit ranges, and considering both a parabolic and a quartic fit function in $\theta_I$. Results marked with * were not considered when assessing the systematic error on $R$ due to the truncation of the Taylor series because of the unreliability of the obtained best fits.}
\label{tab:SU6_comp}
\end{table}

It is worth recalling that, as already explained in Sec.~\ref{sec:res_N4}, the critical temperature $T_c(\theta_L)/\sqrt{\sigma}$ has two sources of uncertainties: the error on $\beta_c(\theta_L)$, and the error on $a(\beta_c(\theta_L))$ due to the scale setting, which were treated separately in our analysis due to the existing correlations among $a(\beta_c(\theta_L))$. Unlike what was found for $N=4$, where the first source of uncertainty gave by far the dominant contribution to the final error on $R$ obtained from the imaginary-$\theta$ fit, in this case the situation is different. Indeed, we estimated from a bootstrap analysis that, for our $N=6$ data, these two sources of uncertainty on $T_c/\sqrt{\sigma}$ gave comparable contributions to the final errors on $R$. Therefore, the uncertainties on $R$ reported in Tab.~\ref{tab:SU6_comp} have been computed from the sum of these two errors, which were added in quadrature.

\begin{table}[!htb]
\begin{tabular}{|c|c|c|c|c|c|c|}
\hline
\multicolumn{7}{|c|}{$N=6$}\\
\hline
\multicolumn{5}{|c|}{Latent heat $\theta=0$ runs} & & \multicolumn{1}{c|}{Imaginary-$\theta$ runs}\\
\cline{1-5}\cline{7-7}
&&&&&&\\[-1em]
$N_s$ & $N_t$ & $\dfrac{\Delta \chi}{T_c^4}$ & $\dfrac{\Delta \epsilon}{T_c^4}$ & $R=\dfrac{\Delta\chi}{2\Delta\epsilon}$ && $R$ (imaginary-$\theta$ fit) \\
&&&&&&\\[-1em]
\cline{1-5}\cline{7-7}
&&&&&&\\[-1em]
10 & 5 & 0.2013(13) & 14.22(43) & 0.00703(22) && 0.00948(50)   \\
12 & 6 & 0.1434(96) & 11.61(21) & 0.00614(37) && 0.00800(50) \\
16 & 8 & 0.1463(46) & 12.17(76) & 0.00608(39) && 0.00598(39) \\
\hline
\end{tabular}
\caption{Summary of our final results for the curvature $R$ for $N=6$ obtained from dedicated $\theta=0$ simulations aiming at determining the jump of the topological susceptibility $\Delta \chi$ and the latent heat $\Delta \epsilon$. For completeness we also report the final results for $R$ obtained from the imaginary-$\theta$ fit.}
\label{tab:FINAL_RES_SU6}
\end{table}

Concerning instead the determination of $R$ from the latent heat, again we relied on dedicated $\theta=0$ simulations performed at $\beta\simeq \beta_c$ and using the same volumes adopted for the identification of the critical point. As already discussed in the previous section, we assigned configurations satisfying suitable cuts on the Polyakov loop to either one of the two phases. Also in this case we verified that changing the values of these cuts within the depleted region between the two peaks has a negligible impact on the latent heat, as any corresponding variation of $\Delta \epsilon$ is much smaller than the obtained statistical errors on this quantity.

\begin{figure}[!htb]
\centering
\includegraphics[width=0.48\textwidth]{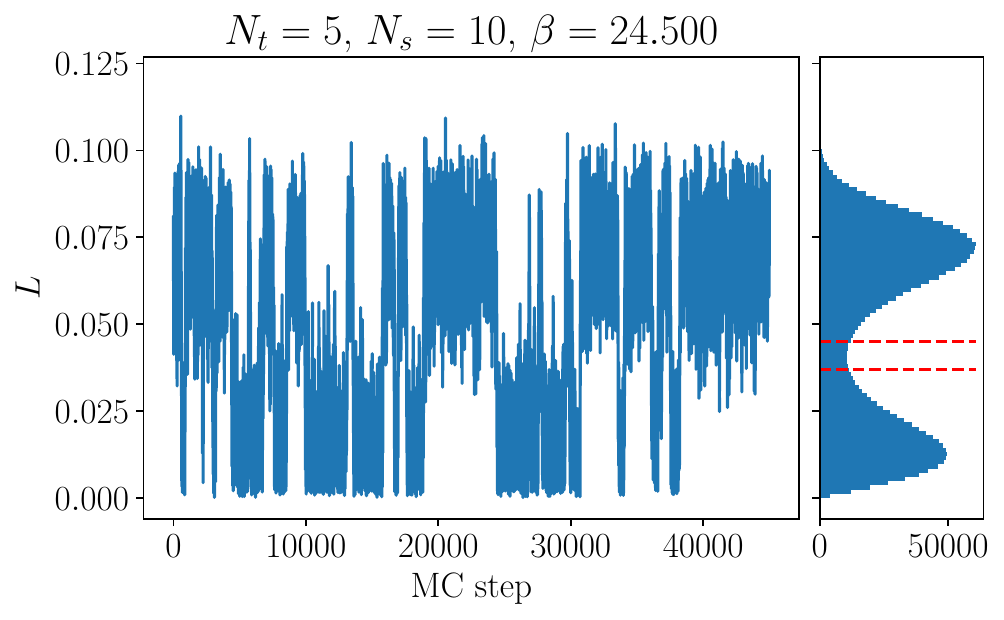}
\caption{History and histogram of the Polyakov loop $L$ for $N=6$ obtained on a $10^3 \times 5$ for $\beta\simeq \beta_c(\theta=0)$. The two dashed lines correspond to the cuts $L_{\mathrm{cut}}^{(c)}$ and $L_{\mathrm{cut}}^{(d)}$ used to assign a configuration to the confined or deconfined phase. The horizontal MC time is expressed in terms of the number of single MC steps defined at the beginning of Sec.~\ref{sec:algos}.}
\label{fig:poly_histo_2peaks_large_vol_N6}
\end{figure}

An example of the histogram of the Polyakov loop sand of the cuts used to separate the two phases is shown in Fig.~\ref{fig:poly_histo_2peaks_large_vol_N6}, while our determinations of $\Delta \chi/T_c^4$, $\Delta \epsilon/T_c^4$ and $R=\Delta \chi/(2 \Delta \epsilon)$ for all explored points are collected in Tab.~\ref{tab:FINAL_RES_SU6}.

\begin{figure}[!htb]
\centering
\includegraphics[width=0.49\textwidth]{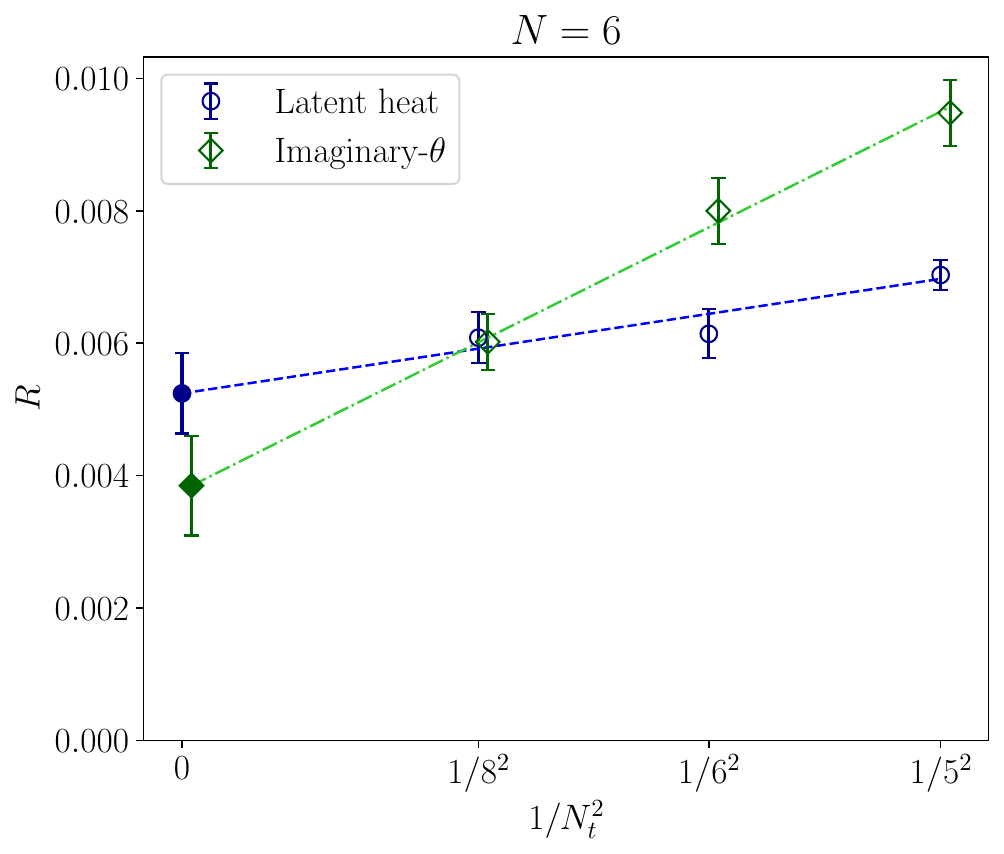}
\includegraphics[width=0.49\textwidth]{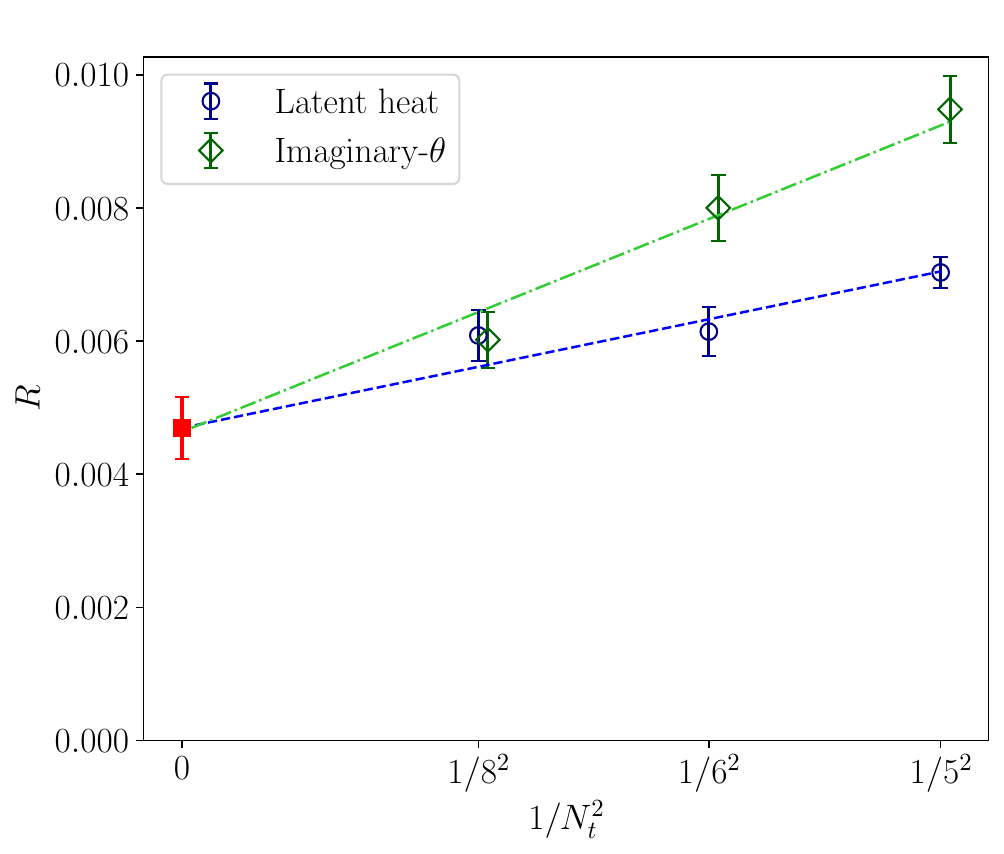}
\caption{Extrapolation towards the continuum limit according to fit function~\eqref{eq:fit_continuum} of the determinations of the curvature $R(N=6)$ obtained both from the imaginary-$\theta$ fit of the critical temperature and from the latent heat of the transition.}
\label{fig:cont_limit_N6}
\end{figure}

The continuum limits of $R(N=6)$ from both data sets, again taken assuming leading $O(a^2)$ corrections, is shown in the top panel of Fig.~\ref{fig:cont_limit_N6}. Our final continuum results are:
\beq
R(N=6) &= 0.00385(75) \,\, &\text{ (from imaginary-$\theta$),}\\
R(N=6) &= 0.00524(61) \,\, &\text{ (from latent heat)}.
\eeq
Again, as it can be observed, obtained determinations for $R$ from the two different strategies are perfectly compatible among them in the continuum limit, confirming again the validity of the prediction in Eq.~\eqref{eq:R_conjecture} for $N=6$ too.

Given the very good agreement between these two continuum extrapolations, also in this case we consider a combined continuum limit of the two data sets, which yields a reduced chi-square of 3.18/3, corresponding to a $p$-value of $37\%$, and a combined limit:
\beq\label{eq:cont_lim_combo_N6}
R(N=6) = 0.00469(47)\quad \text{ (combined)}.
\eeq
The combined limit is shown in the bottom panel of Fig.~\ref{fig:cont_limit_N6}.

\FloatBarrier

\subsection{The large-$N$ limit}

This section is devoted to the discussion of the large-$N$ limit of $R$, which we extrapolate using our determinations for $N=4,6$ and the pre-existing $N=3$ results. Our goal is in particular to compare the large-$N$ scaling obtained from the imaginary-$\theta$ and the latent heat determinations.

The results for $R$ as a function of $N$ are summarized in Tab.~\ref{tab:FINAL_RES_TOT}. Concerning the $N=3$ determinations, we considered the value of $R(N=3)$ obtained from the imaginary-$\theta$ fit of Ref.~\cite{DElia:2012pvq}, and the values of $\Delta \epsilon(N=3)$ and $\Delta \chi(N=3)$ obtained, respectively, in Refs.~\cite{Borsanyi:2022xml} and~\cite{Borsanyi:2022fub} to compute $R=\Delta \chi / (2 \Delta \epsilon)$ for $N=3$.

\begin{table}[!htb]
\begin{tabular}{|c|c|c|}
\hline
$N$ & $R$ (Imaginary-$\theta$) & $R=\Delta \chi/(2 \Delta \epsilon)$ \\
\hline
3 & 0.0178(5)   & 0.0168(27) \\
4 & 0.0095(11)  & 0.0119(16) \\
6 & 0.00385(75) & 0.00524(61) \\
\hline
\end{tabular}
\caption{Summary of our final results for the curvature $R$ for $N=4$ and $6$, along with the $N=3$ results obtained in Refs.~\cite{DElia:2013uaf} (imaginary-$\theta$) and~\cite{Borsanyi:2022xml,Borsanyi:2022fub} (latent heat and jump of the susceptibility).}
\label{tab:FINAL_RES_TOT}
\end{table}

As explained earlier in the introduction, we expect the following asymptotic large-$N$ scaling:
\beq
R = \frac{\bar{R}}{N^2} + O\left(\frac{1}{N^4}\right).
\eeq
In order to verify this prediction, we considered the following 3 ans\"atze, which we used to perform a best fit of the two data sets reported in Tab.~\ref{tab:FINAL_RES_TOT}:
\beq
\label{eq:largeN_ansatz1}
R(N) &=& \frac{\bar{R}}{N^c},
\eeq

\beq
\label{eq:largeN_ansatz2}
R(N) &=& \frac{\bar{R}}{N^2},
\eeq

\beq
\label{eq:largeN_ansatz3}
R(N) &=& \frac{\bar{R}}{N^2} + \frac{\bar{R}^{(1)}}{N^4}.
\eeq

We start our investigation by comparing the best fits obtained from these ans\"atze using the imaginary-$\theta$ data set. Assuming a pure power-law behavior $R\sim 1/N^c$, with $c$ left as a free parameter, the best fit of $R(N)$ as a function of $1/N$ yields $c = 2.20(24)$, which is in perfect agreement with the $1/N^2$ predicted leading large-$N$ scaling for $R$. This finding thus justifies a best fit fixing $c=2$, which gives a perfectly compatible results for the pre-factor $\bar{R}$, both when the determination for $N=3$ is included and excluded from the fit. Finally, fitting our data assuming the presence of further $1/N^4$ higher-order corrections also gives perfectly agreeing results with the pure $1/N^2$ fit, as we find a compatible result for $\bar{R}$, and a value for the $O(1/N^4)$ coefficient $\bar{R}^{(1)} = 0.22(25)$ which is compatible with zero within the error.

\begin{figure}[!t]
\centering
\includegraphics[width=0.48\textwidth]{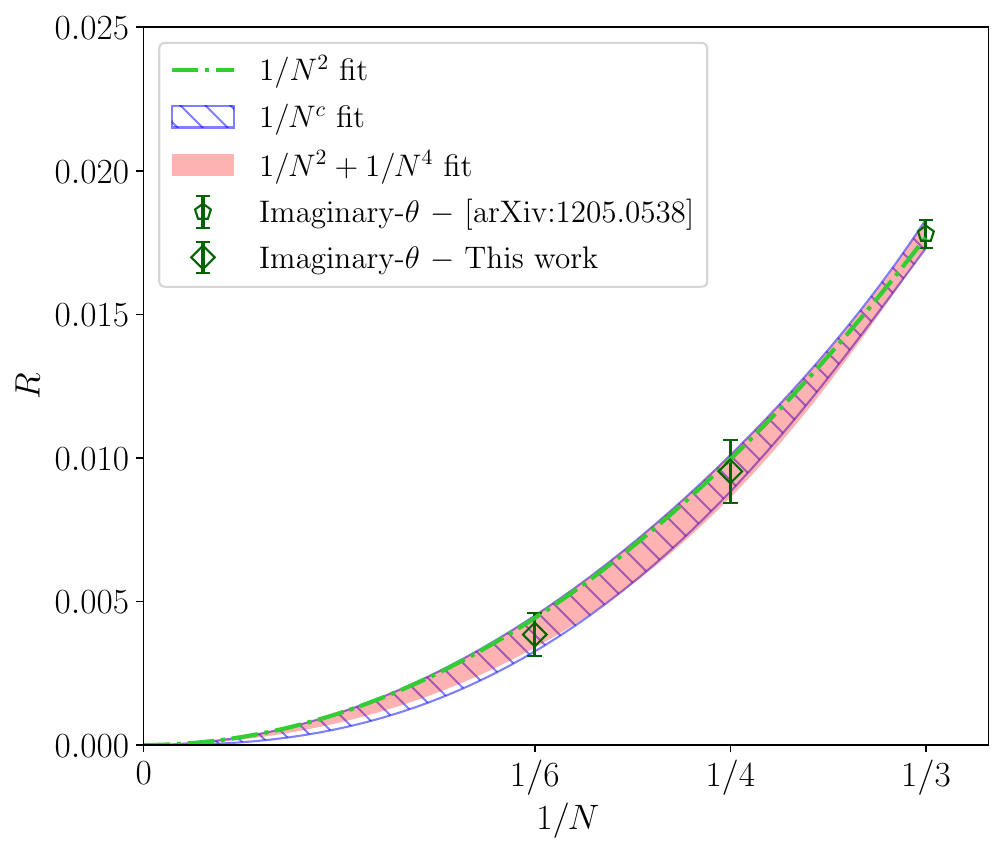}
\includegraphics[width=0.48\textwidth]{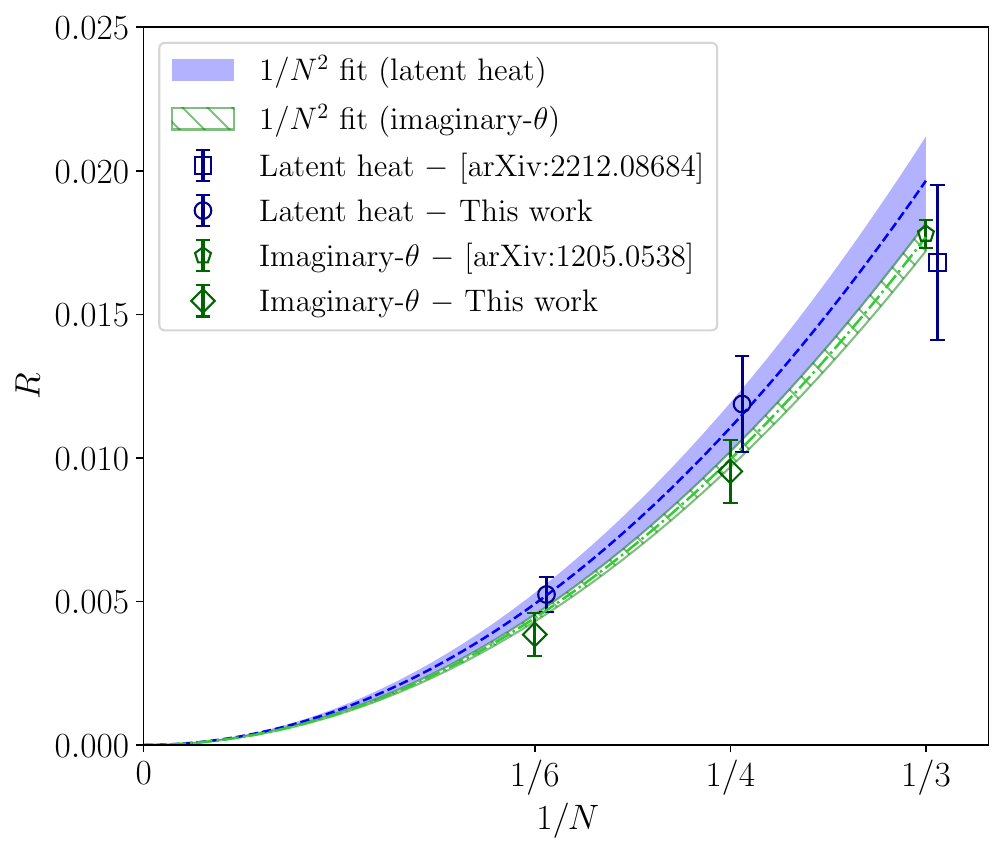}
\caption{Study of the large-$N$ scaling of $R$. Top panel: Solid, dashed and dotted lines represent, respectively, the best fit of our data for $R(N)$ in Tab.~\ref{tab:FINAL_RES_TOT} according to fit functions~\eqref{eq:largeN_ansatz1},~\eqref{eq:largeN_ansatz2} and~\eqref{eq:largeN_ansatz3}. Bottom panel: comparison of the large-$N$ scaling of the determinations of $R(N)$ obtained from the imaginary-$\theta$ fit and from the latent heat. Dashed and dot-dashed lines represent the best fit of both data sets according to fit function~\eqref{eq:largeN_ansatz2}.}
\label{fig:fit_largeN}
\end{figure}

\begin{figure}[!t]
\centering
\includegraphics[width=0.48\textwidth]{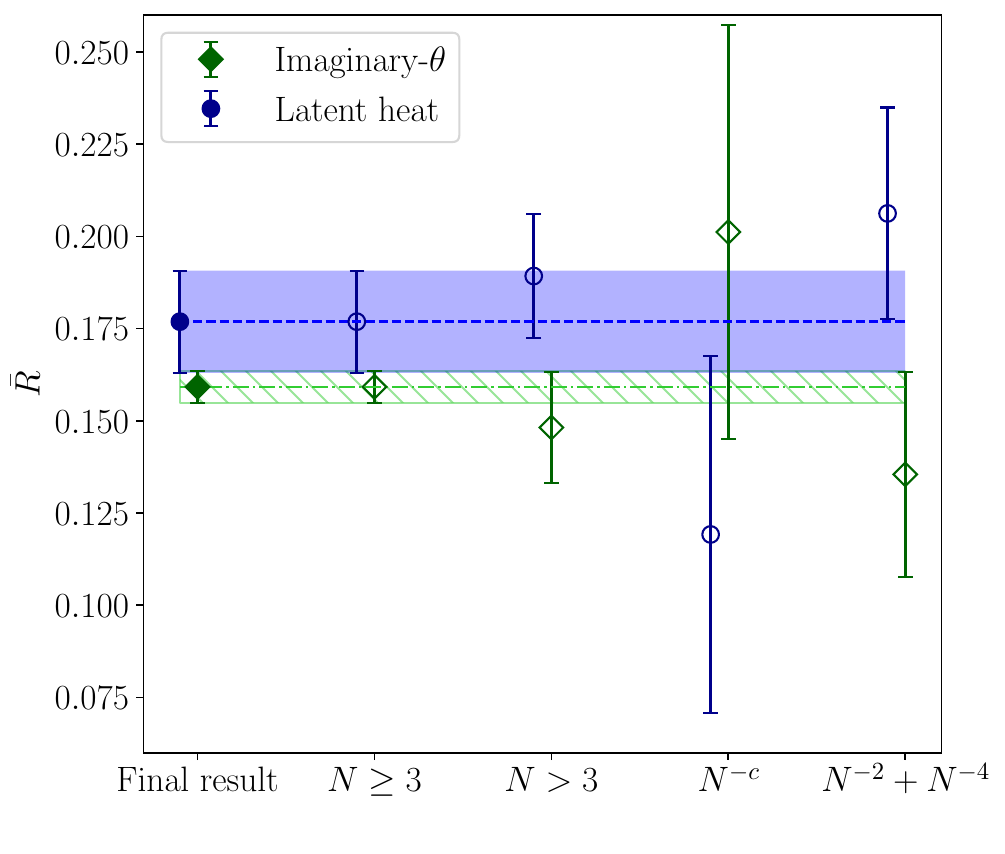}
\caption{Comparison of the values obtained for the pre-factor of the $O(1/N^2)$ coefficient of $R(N)$ using fit functions~\eqref{eq:largeN_ansatz1},~\eqref{eq:largeN_ansatz2} and~\eqref{eq:largeN_ansatz3} to fit our results for $R(N)$ in Tab.~\ref{tab:FINAL_RES_TOT}.}
\label{fig:syst_Rbar}
\end{figure}

The perfect agreement among the best fitting curves obtained from the 3 different ans\"atze employed can also be clearly seen from the top panel of Fig.~\ref{fig:fit_largeN}, where one can clearly observe the 3 curves and their corresponding error band to be perfectly overlapping bands, and also from the comparison reported in Fig.~\ref{fig:syst_Rbar} of the different values of $\bar{R}$ obtained from the various employed ans\"atze, which are all in very good agreement.

Perfectly agreeing results and perfectly analogous considerations can be drawn by repeating this systematic study of the $N$-dependence of $R$ from the latent heat data set, as it can be seen from the comparison in the bottom panel of Fig.~\ref{fig:fit_largeN}, and from the obtained results for $\bar{R}$ from these determinations reported in Fig.~\ref{fig:syst_Rbar}.

As a matter of fact, from the comparison of obtained results for $\bar{R}$ from the various employed ans\"atze, cf.~Fig.~\ref{fig:syst_Rbar}, we quote as our final results:
\beq\label{eq:final_res_Rbar}
\begin{aligned}
R &\simeq\frac{\bar{R}}{N^2},& \quad &N\ge 3,\\
\bar{R} &= 0.159(4),&  \quad &(\text{from imaginary-$\theta$}),\\
\bar{R} &= 0.177(14),& \quad &(\text{from latent heat}).
\end{aligned}
\eeq

Summarizing, we can conclude that our numerical results for $N=4,6$, along with the previously-obtained $N=3$ determinations, show that $R(N)$ follows the predicted leading large-$N$ scaling, thus confirming the conjectured relation between the curvature of the phase diagram in the $T-\theta$ plane and the topological features of the $\theta=0$ critical deconfinement transition.

One intriguing outcome of our results regarding the leading large-$N$ scaling of $R$ is the confirmation that $\Delta \chi = \chi_c - \chi_d \sim O(N^0)$, as expected on the basis of general theoretical arguments, since $\chi_c \sim O(N^0)$, according to the Witten--Veneziano mechanism, and $\chi_d \sim O(T_c^{-N})$, according to the DIGA prediction. However, while the former is a very well established fact from lattice simulations~\cite{Ce:2016awn,Bonati:2016tvi,Bonanno:2020hht,Athenodorou:2022aay}, the former has not been directly verified, even though there is numerical evidence that $\chi_d/\chi_c = \chi(T_c^+)/\chi(T_c^-)$ is rapidly suppressed as a function of $N$~\cite{Lucini:2004yh,DelDebbio:2004vxo}. Since to obtain $\Delta \chi$ we computed $\chi_d = \chi(T_c^{+})$ at $\theta=0$, a byproduct of our large-$N$ investigation of $R$ is the first direct lattice study of the $N$-dependence of $\chi_d$, which we are able to perform combining our continuum results for $N=4$ and $6$ with those obtained for $N=3$ in Ref.~\cite{Borsanyi:2022xml}.

\begin{figure*}[!t]
\centering
\includegraphics[width=0.58\textwidth]{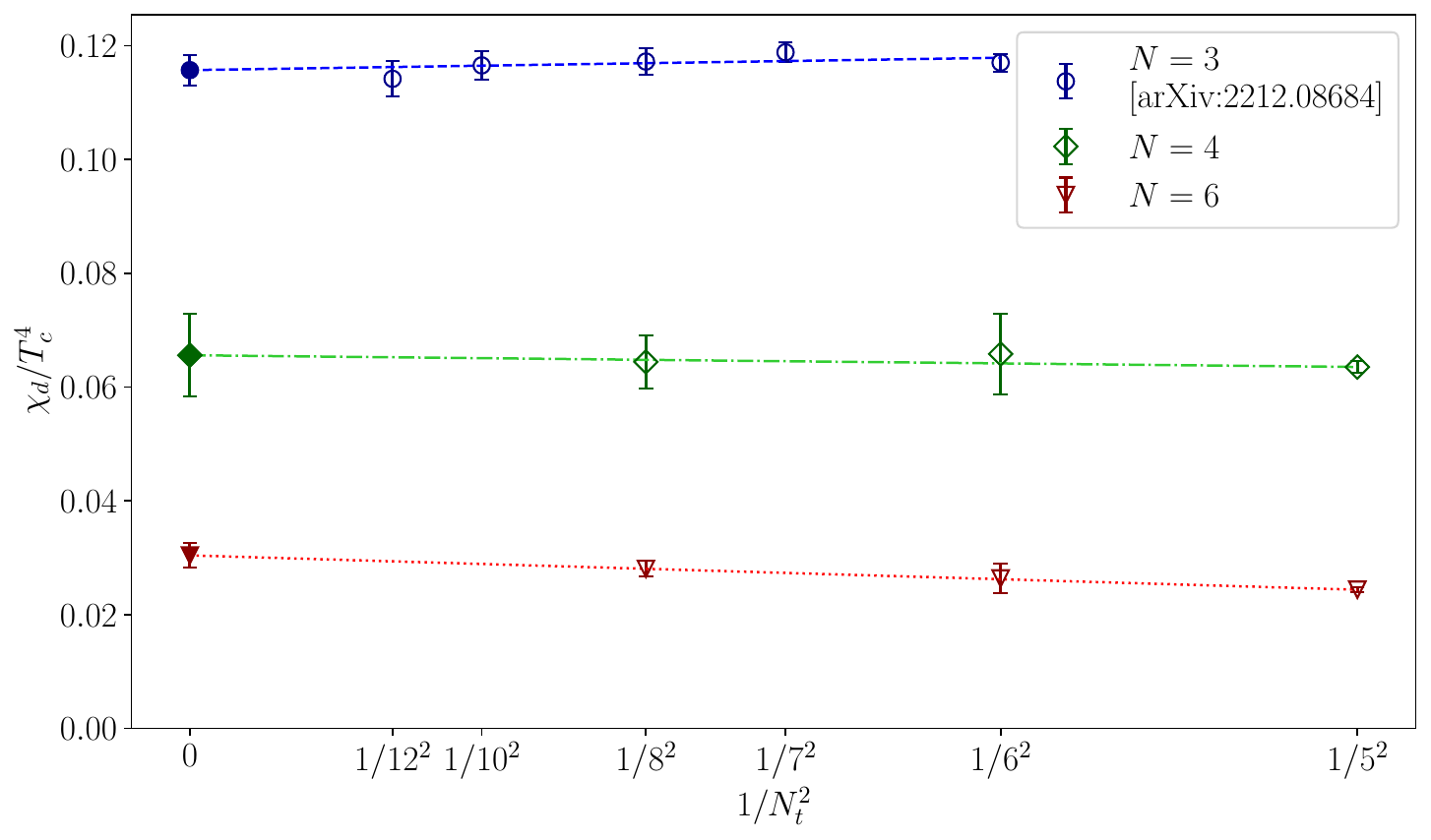}
\includegraphics[width=0.4\textwidth]{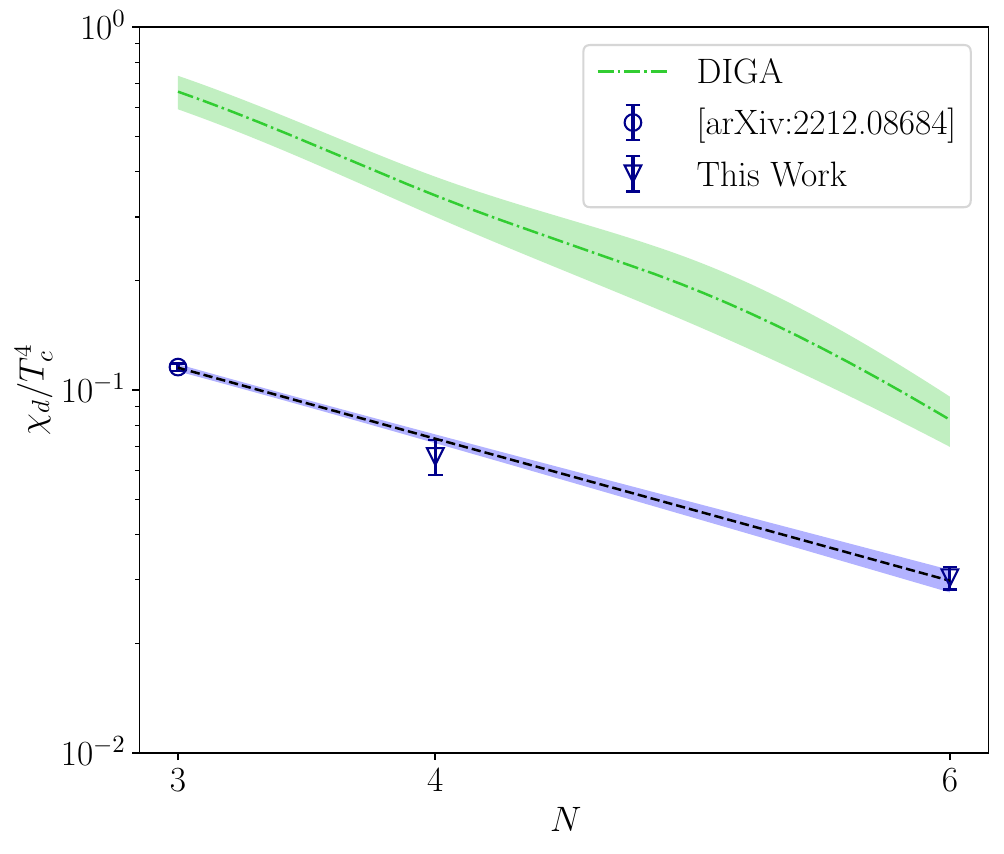}
\caption{Left panel: continuum extrapolations of our determinations of $\chi_d/T_c^4=\chi(T_c^{+})/T_c^4$ for $N=4$ and $6$, along with the extrapolation of the $N=3$ ones found in Tab.~2 of Ref.~\cite{Borsanyi:2022fub}. Right panel: behavior of the continuum-extrapolated determinations of $\chi_d/T_c^4$ as a function of $N$. The dashed line represents an exponential best fit of $\chi_d/T_c^4$ as a function of $N$, as predicted by the DIGA.}
\label{fig:chi_d_largeN}
\end{figure*}

The extrapolation towards the continuum limit of our values of $\chi_d/T_c^4$ for $\SU(4)$ and $\SU(6)$, as well as that of the $\SU(3)$ results reported in Tab.~2 of Ref.~\cite{Borsanyi:2022fub} are shown in the left panel of Fig.~\ref{fig:chi_d_largeN}. The continuum limits of $\chi_d/T_c^4$ for these 3 values of $N$ are reported in Tab.~\ref{tab:chi_d_continuum}. As it can be observed, they can be nicely fitted assuming an exponential suppression in $N$, as predicted by the DIGA. The exponential best fit of the continuum results for $\chi_d$ as a function of $N$, shown in the right panel of Fig.~\ref{fig:chi_d_largeN}, performed according to the fit function $A\exp(-b N)$, has a reduced chi-squared $\tilde{\chi}^2=1.27/1$, leading to a p-value of $\sim 74\%$, thus providing an excellent description of these data. The best fit parameters turn out to be $A = 0.446(38)$ and $b=0.451(25)$.

The one-loop DIGA prediction, however, overestimates $\chi_d/T_c^4$ by almost an order of magnitude, as it can be observed from the comparison in Fig.~\ref{fig:chi_d_largeN}. On the other hand, the decay width of the exponential, although not exactly compatible, is in qualitative good agreement with lattice results. As a matter of fact, performing an exponential best of the numerical DIGA prediction according to $A_{\rm DIGA}\exp(-b_{\rm DIGA} N)$ yields $A_{\rm DIGA} = 4.7(3)$ and $b_{\rm DIGA} = 0.65(2)$ (technical details about the numerical calculation of the DIGA prediction can be found in App.~\ref{app:DIGA}). This is not surprising, as similar conclusions have been reached by comparing the lattice results for the temperature dependence of the topological susceptibility in the deconfined phase with the DIGA prediction both in quenched QCD and in the full theory with dynamical fermions~\cite{Borsanyi:2015cka,Petreczky:2016vrs,Frison:2016vuc,Borsanyi:2016ksw,Borsanyi:2021gqg,Athenodorou:2022aay} \footnote{One-loop DIGA computations are very sensitive to the choice of renormalization scheme, and thus to the value of the coupling. For example, in Ref.~\cite{Borsanyi:2015cka} one-loop DIGA result for $\chi(N=3)$ at $T=T_c$ were found to \emph{underestimate} the lattice one by an order of magnitude. The origin of this discrepancy can be traced back to the fact that the authors employed the four-loop renormalized coupling, while here we considered just the one-loop expression.}.

\begin{table}[!t]
\centering
\begin{tabular}{|c|c|}
\hline
$N$ & $\chi_d / T_c^4 = \chi(T_c^{+})/T_c^4$ \\
\hline
3 & 0.1157(27) \\
4 & 0.0656(73) \\
6 & 0.0304(22) \\
\hline
\end{tabular}
\caption{Continuum determinations of the topological susceptibility obtained at $T=T_c$ in the deconfined phase, $\chi_d/T_c^4=\chi(T_c^+)/T_c^4$, as a function of $N$. The result for $N=3$ was obtained extrapolating towards the continuum limit the determinations in Tab.~2 of Ref.~\cite{Borsanyi:2022fub}.}
\label{tab:chi_d_continuum}
\end{table}

\section{Conclusions}\label{sec:conclu}

In this work we have presented a large-$N$ lattice investigation of the $O(\theta^2)$ coefficient $R$, parametrizing the $\theta$-dependence of the critical deconfinement transition of $\SU(N)$ gauge theories at leading order:
\beq
\frac{T_c(\theta)}{T_c(\theta=0)} = 1 - R \theta^2 + O(\theta^4).
\eeq
This quantity can be related to two features of the deconfinement transition at $\theta=0$, namely, the latent heat and the jump of the topological susceptibility, $R=\Delta \chi / (2 \Delta \epsilon)$, implying also $R \sim O(1/N^2)$.

In order to verify this prediction, we performed a direct lattice calculation of $R$ for $N=4$ and $6$ using two different and complementary strategies. The first one relied on the fit of the imaginary-$\theta$ dependence of the critical temperature, which was achieved by performing simulations in the presence of a non-vanishing imaginary-$\theta$. The second strategy instead is based on the calculation for vanishing $\theta$ of the latent heat and of the jump of $\chi$ at the critical point, and thus relied just on $\theta=0$ simulations.

In both cases, for the finest lattice spacings explored at $N=6$, due to the significant topological freezing suffered from standard simulations, we adopted the recently-proposed state-of-the-art PTBC algorithm to obtain a sizable reduction of the auto-correlation time of the topological charge.

Our numerical results show that, after the continuum limit is taken, the two different determinations of $R$ perfectly agree among each other, thus confirming the predicted relation between $R$, $\Delta \epsilon$ and $\Delta \chi$ for several values of $N$. Moreover, combining our new $N=4,6$ results with the pre-existing $N=3$ determination of $R$, $\Delta \chi$ and $\Delta \epsilon$, we were able to study the large-$N$ behavior of $R(N)$, finding perfect agreeing results with the predicted leading $1/N^2$ scaling already for $N\ge 3$. As a byproduct of this investigation, we were also able to show that the topological susceptibility in the deconfined phase computed at the critical temperature is exponentially suppressed as a function of $N$, as predicted by the DIGA. While the DIGA decay width qualitatively agrees with lattice data, the DIGA pre-factor of the exponential overestimates the lattice result by an order of magnitude.

In conclusion, we quote as our final result:
\beq
R(N) \simeq \frac{0.159(4)}{N^2}, \quad N\ge 3.
\\\nonumber
\eeq

This result fits very well with the picture emerged from several other previous large-$N$ studies of $\SU(N)$ pure-gauge theories, both done using regular extended lattices and volume-reduced models, pointing out a rapid approach of pure Yang--Mills theories towards the large-$N$ limit~\cite{Lucini:2012gg,Bali:2013kia,Bonati:2016tvi,Ce:2016awn,GarciaPerez:2020gnf,Hernandez:2020tbc,Perez:2020vbn,Bonanno:2020hht,Athenodorou:2021qvs,Bonanno:2023ypf,Bonanno:2023ple}.

\acknowledgements
It is a pleasure to thank B.~All\'es, C.~Bonati and E.~Vicari for useful discussions. We also would like to thank the anonymous referee for suggesting to include the study of the $N$-dependence of $\chi_d$. The work of C.~B.~is supported by the Spanish Research Agency (Agencia Estatal de Investigación) through the grant IFT Centro de Excelencia Severo Ochoa CEX2020-001007-S and, partially, by grant PID2021-127526NB-I00, both funded by MCIN/AEI/10.13039/501100011033. This work has been partially supported by the project "Non-perturbative aspects of fundamental interactions, in the Standard Model and beyond" funded by MUR,  Progetti di Ricerca di Rilevante Interesse Nazionale (PRIN), Bando 2022, grant 2022TJFCYB (CUP I53D23001440006). Numerical simulations have been performed on the \texttt{MARCONI} machine at Cineca, based on the agreement between INFN and Cineca, under projects INF22\_npqcd and INF23\_npqcd.

\appendix

\section*{Appendix}
\section{One-loop DIGA prediction for $\chi(T_c)$ in the deconfined phase}\label{app:DIGA}

Using the semiclassical expansion of the QCD path-integral and perturbation theory around the 1-instanton sector, one obtains the following DIGA prediction~\cite{RevModPhys.53.43,Boccaletti:2020mxu}:
\beq\label{eq:DIGA_chi}
    \chi(T)\Big\vert_{\rm DIGA} \simeq 2 \int_0^{\infty} n(\rho) e^{-S(\rho,T)} \, d\rho, \quad T \gg T_c.
\eeq
The instanton number density is given by~\cite{RevModPhys.53.43,Boccaletti:2020mxu}:
\begin{align}
\nonumber
n(\rho) &= C \mu^5 \left(\frac{16\pi^2}{g^2(\mu)}\right)^{2N} e^{-\frac{8\pi^2}{g^2(\mu)}} (\rho \mu)^{\beta_1-5} \prod_{f=1}^{N_f}[\rho m_f(\mu)],\\
\beta_1 &= \frac{11 N-2N_f}{3}.    
\end{align}
The pre-factor $C$ appearing in $n(\rho)$ is scheme-dependent; here we choose the $\overline{\mathrm{MS}}$ scheme:
\beq
\begin{aligned}
C_{\overline{\mathrm{MS}}} &= \frac{e^{c_0+c_1 N+c_2 N_f}}{(N-1)!(N-2)!},\\
c_0 = -0.76297926, \,\, c_1 &= -2.89766868, \,\, c_2 =  0.26144360.
\end{aligned}
\eeq

The renormalized running coupling $g^2(\mu)$ appearing in $n(\rho)$ is instead given at one-loop in the $\overline{\mathrm{MS}}$ scheme by~\cite{Steffens:2004sg}:
\beq
\label{eq:oneloop_coupling}
\frac{g^2(\mu)}{8 \pi^2} = \frac{1}{\beta_1\log\left(\mu/\Lambda_{\overline{\mathrm{MS}}}\right)}
\eeq
where $\Lambda_{\overline{\mathrm{MS}}}$ is the dynamically-generated scale, and where the renormalization scale $\mu$ is taken to be $\mu = c T$ with $c>1$ some constant (the dependence on $c$ is indicative of higher-order corrections). In the following we will use $\mu=\pi T$. The quantity $m_f(\mu)$ is instead the running renormalized quark mass with flavor $f$, which will not be considered here as we will focus on the quenched case $N_f=0$.

The function $S(\rho,T)$ is actually just a function of the variable $\lambda \equiv \pi \rho T$, which explains our choice $\mu=\pi T$ for the renormalization scale:
\beq
S(\lambda) = \frac{\lambda^2}{3}(2N+N_f) + 12A(\lambda)\left(1+\frac{N-N_f}{6}\right) \quad\qquad
\eeq

The function $A(\lambda)$ is the finite part resulting from the subtraction of two divergent integrals. An exact close form is not known; however, it can be computed numerically, and asymptotic analytic expressions for $\lambda\to 0$ and $\lambda\to\infty$ are known.  In~\cite{Boccaletti:2020mxu}, however, a practical approximate parameterization is provided, which will be employed here:
\beqnn
-12 A_{\rm param}(\lambda) \approx p_0 \log(1 + p_1\lambda^2 + p_2\lambda^4 + p_3 \lambda^6 + p_4 \lambda^8),
\eeqnn
with
\begin{align}
\nonumber
p_0 &= 0.247153244, & p_1 &= 1.356391323,\\
\nonumber
p_2 &= 0.675021523, & p_3 &= 0.145446632,\\
p_4 &= 0.008359667. &
\end{align}
This parameterization approximates the numerically-computed $A(\lambda)$ with an $\sim O(10^{-4})$ accuracy in the range $\lambda\in[0,10]$, which is largely sufficient to compute~\eqref{eq:DIGA_chi}, given the exponential suppression of the integrand.

Putting all together, we find, for the quenched theory ($N_f=0$), the following one-loop $N$-dependence for the topological susceptibility at the critical temperature in the deconfined phase in the $\overline{\mathrm{MS}}$ scheme:
\begin{widetext}
\beq\label{eq:final_chid}
\begin{aligned}
\frac{\chi(T_c^+)}{T_c^4}\Bigg\vert_{\rm DIGA}
=& 2 \pi^{4} C_{\overline{\mathrm{MS}}}^{(N_f=0)} \left[\frac{16\pi^2}{g_{N_f=0}^2(\pi T_c)}\right]^{2N} e^{-\frac{8\pi^2}{g_{N_f=0}^2(\pi T_c)}}
\int_{0}^{\infty} \frac{e^{-S^{(N_f=0)}(\lambda)}}{\lambda^{5-\frac{11N}{3}}} \, d\lambda =\\
=& 2 \pi^{4} C_{\overline{\mathrm{MS}}}^{(N_f=0)} \left[\frac{22N}{3}\log\left(\frac{\pi T_c}{\Lambda_{\overline{\mathrm{MS}}}}\right)\right]^{2N}
\left(\frac{\pi T_c}{\Lambda_{\overline{\mathrm{MS}}}}\right)^{-\frac{11N}{3}} \int_{0}^{\infty} \frac{e^{-S^{(N_f=0)}(\lambda)}}{\lambda^{5-\frac{11N}{3}}} \, d\lambda.
\end{aligned}
\eeq
\end{widetext}
For the purpose of evaluating Eq.~\eqref{eq:final_chid}, the numerical values for $\Lambda_{\overline{\mathrm{MS}}}(N)$ and $T_c(N)$ were taken respectively from Ref.~\cite{Athenodorou:2022aay} and Ref.~\cite{Lucini:2012gg}.

\end{document}